\documentclass[twocolumn]{aastex631}

\usepackage[T1, T2A]{fontenc}
\usepackage[utf8]{inputenc}

\shorttitle{A study of nine extreme low mass ratio contact binary systems}
\shortauthors{E. Lalounta et al.}
\graphicspath{{./}{figures/}}
\begin{document}

\title{A study of nine extreme low mass ratio contact binary systems}

\author{Eleni Lalounta}
\affiliation{Department of Physics, University of Patras, 26500, Patra, Greece}

\correspondingauthor{Panagiota-Eleftheria Christopoulou}
\email{pechris@upatras.gr}

\author{Panagiota-Eleftheria Christopoulou}
\affiliation{Department of Physics, University of Patras, 26500, Patra, Greece}

\author[0000-0002-3039-9257]{Athanasios Papageorgiou}
\affiliation{Department of Physics, University of Patras, 26500, Patra, Greece}
\author{C. E. Ferreira Lopes}
\affiliation{Instituto de Astronomía y Ciencias Planetarias, Universidad de
Atacama, Copayapu 485, Copiap\'{o}, Chile}
\affiliation{Universidade de S\~{a}o Paulo, IAG, Rua do Mat\~{a}o 1226, Cidade Universit\'{a}ria, S\~{a}o Paulo, 05508-900, Brazil}
\affiliation{National Institute For Space Research (INPE/MCTI), Av. dos Astronautas, 1758 - S\~{a}o Jos\'{e} dos Campos - SP, 12227-010, Brazil}

%%\collaboration{6}{(AAS Journals Data Editors)}

\author[0000-0001-6003-8877]{M\'{a}rcio Catelan}
\affiliation{Instituto de Astrofísica, Pontificia Universidad Cat\'{o}lica de Chile,\\
Av. Vicu\~{n}a Mackenna 4860, 7820436 Macul, Santiago, Chile}
\affiliation{Millennium Institute of Astrophysics, Nuncio Monse\~{n}or Sotero Sanz 100,\\ Of. 104, Providencia, Santiago, Chile}
\affiliation{Centro de Astro-Ingeniería, Pontificia Universidad Católica de Chile, Av. Vicuña Mackenna 4860, 7820436 Macul, Santiago, Chile}

%%\author{ }
%%\altaffiliation{ }
%%\affiliation{TeXnology Inc.}

%%\author{Julie Steffen}
%%\affiliation{AAS Director of Publishing}
%%\affiliation{American Astronomical Society \\
%%1667 K Street NW, Suite 800 \\
%%Washington, DC 20006, USA}

%%\author{Magaret Donnelly}
%%\affiliation{IOP Publishing, Washington, DC 20005}



\begin{abstract}
Low mass ratio systems (LMR) are a very interesting class of contact eclipsing binaries challenging the theoretical models of stability. These systems are also considered possible progenitors of the rare low-mass optical transients called red novae. In this study, we present the identification of 7 new totally eclipsing LMR systems from Catalina Sky Surveys (CSS) and 77 LMR candidates from the All Sky Automated Survey (ASAS-3). Using the available CSS light curves and new multiband observations for CSS$\_$J210228.3-031048 and CSS$\_$J231513.3+345335 with the 2.3 m Aristarchos telescope at Helmos Observatory, we estimate their physical and absolute parameters and investigate their stability and their progenitors. The light curves are analyzed by performing a 2-dimension scan on the mass ratio – inclination plane with Phoebe-0.31 scripter while the errors are estimated using Monte-Carlo simulations and heuristic scanning of the parameter space. Our analysis revealed that all 9 CSS systems have extreme mass ratios from 0.09 to 0.16. 
Our statistical analysis of well-studied LMR contact binaries shows that LMR systems tend to have warmer and more massive primaries. The investigation of the progenitors of both low and higher-mass ratio systems reveals a trend for the former to originate from higher-mass ancestors. Finally, we investigate the stability condition by calculating the ratio of spin angular momentum to orbital angular momentum and other stability indicators in the context of the reliability of the solutions.

\end{abstract}

\keywords{Surveys - binaries: eclipsing - Stars: fundamental parameters - Stars: evolution}


\section{Introduction} \label{sec:intro}
The discovery of a plethora of binaries in the era of large surveys, the availability of advanced follow-up facilities, and the analysis of observations with sophisticated modeling tools have recently led to a resurgence in the study of the origin, structure, and evolution of binary systems.

Contact binaries (also called EW or W UMa systems) are composed of two Roche lobe-filling, strongly interacting stars, usually of late type, that share a common envelope. Detached cool binaries with an initial period of a few days that lose mass and angular momentum through magnetic braking are thought to be their progenitors. The time evolution of the orbital period and observed mass ratio of EWs are governed by the interplay between mass transfer/loss processes, angular momentum loss (AML), tidal effects, and the presence of additional components. How contact binaries of low/moderate mass end their lives catastrophically by merging into a single star or going through ``common envelope'' events is still debated. Although theoretical models of an EW containing two main sequence stars predict coalescence due to the tidal Darwin instability \citep[e.g.,][]{10.2307/113751, 1980A&A....92..167H} at a minimum secondary-to-primary mass ratio ($q_{\rm min}$) below 0.07-0.1 \citep{1995ApJ...444L..41R, 2009PASP..121.1036A}, many new EWs beyond this limit have been discovered photometrically \citep[][and references therein]{2022MNRAS.512.1244C} by \textit{Kepler} and ground-based wide-field surveys, including the Robotic Optical Transient Search Experiment \citep[ROTSE,] [] {2000AJ....119.1901A}, All Sky Automated Survey \citep[ASAS,][]{ASAS-3,2002AcA....52..397P}, Super Wide Angle Search for Planets \citep[SuperWASP,][]{2010A&A...520L..10B}, Catalina Sky Surveys  \citep[CSS,][]{2009ApJ...696..870D, 2014ApJS..213....9D}, Asteroid Terrestrial-impact Last Alert System \citep[ATLAS,][]{2018AJ....156..241H} and Zwicky Transient Facility \citep[ZTF,][]{2019PASP..131a8002B}. This notwithstanding, there is only one documented merging event in the Milky Way, V1309 Sco \citep{2011A&A...528A.114T}, which became the prototype of a new class of low mass optical transients known as red novae (RN). According to observational pre-outburst constraints, retrieved from archival data, many researchers \citep{2011A&A...531A..18S,2014ApJ...786...39N} have shown that V1309 Sco's progenitor was an eclipsing contact binary with a period of 1.4 days and an extremely low mass ratio, composed of an early subgiant and a low-mass companion. \cite{2014MNRAS.443.1319K} estimated that Galactic stellar mergers such as V1309 Sco occur once per decade. Therefore, the identification of candidate systems before the merger event among low-mass ratio (LMR) contact systems is of great importance for the understanding of the mechanisms involved.

The suitability of survey data for light curve analysis has not been thoroughly examined. According to \cite{2020ApJS..247...50S}, there was good agreement between ASAS-SN photometric data in comparison to ground-based observations regarding the mass ratio. This was also confirmed by \cite{2022MNRAS.512.1244C} between CSS and dedicated data. \cite{2020MNRAS.493.1565D}, \cite{2021MNRAS.506.4251Z} and  \cite{2022JApA...43...94W} showed the same agreement on the mass ratio and fractional radii, with some variations in the fillout factor and temperature ratio.

As part of an ongoing project to find LMRs in CSS data, we present the identification and photometric investigation of 7 new totally eclipsing LMRs found in CSS. Additionally, we conduct multiband photometry and analysis for two totally eclipsing LMRs \citep{2020CoSka..50..409L} included also in the automatic photometric analysis of \cite{2020ApJS..247...50S} of 2335 late-type contact binaries from CSS. 

The 9 systems were first classified as EW-type by \cite{2014MNRAS.441.1186D}. Later, \cite{2020CoSka..50..409L} included these systems in a sample of 2101 LMR candidates showing total eclipses found in CSS, applying Fourier decomposition on the phase-folded and normalized flux light curves (LCs). Recently, \cite{2022MNRAS.512.1244C} presented the identification and photometric analysis of the first 30 new LMRs from the above sample (see the paper for details of the methodology employed).
   
The paper is organized as follows: in Sect.~\ref{sec:Observations}, we present all the information about the observations of the objects together with  CSS data. In Sect.~\ref{sect:modelling}, we describe the photometric analysis of the light curves, and the physical and absolute parameters determination with their errors. Finally, in Sect.~\ref{sec:Discussion} we discuss our results and investigate the possibility of our sources being merger candidates. New possible LMRs from ASAS-3 are also presented in the Appendix~\ref{sub:ASAS-3}.
\section{CSS and new photometric observations} \label{sec:Observations}
In this study, we use the photometric data obtained from the Catalina Real-Time Transient Survey Data Release 2 \citep[CRTS DR2;][]{2014MNRAS.441.1186D}, which covers 9 years (2004-2013). The observations are taken unfiltered to maximize the throughput and then the magnitudes are transformed to an approximate V magnitude \citep[V$_{\rm CSS}$,][]{2013ApJ...763...32D}. The 9 studied systems have sufficiently sampled LCs with more than 200 observations ($N_{\rm p}$). Table~\ref{tab:1} provides the basic information of the 9 LMR EWs, such as the coordinates (${\rm RA}_{J2000}$ and ${\rm DEC}_{J2000}$), the reference time of minimum (in Heliocentric Julian Days, $\rm HJD_0$), Period (in days), the CSS magnitude at maximum light ($V_{\rm CSS}({\rm max})$), the mean photometric error \citep[$\sigma_{\rm CSS}$, calculated according to the analytical expression derived by][their footnote~8]{2018ApJS..238....4P}, and the number of available CSS observations ($N_{\rm p}$).

In addition, follow-up multicolor observations were carried out for the two shortest-period of the 9 systems, J210228 (0.2455 d) and J231513 (0.2661 d) using the Aristarchos telescope at Helmos Observatory, Greece. This 2.3\,m Ritchey-Chr\'{e}tien telescope is equipped with a liquid nitrogen-cooled VersArray 1024B CCD camera with 1024 $\times$ 1024 pixels and a standard $BVRI$ filter set. The pixel scale is 0.28$\arcsec$, which gives an effective field of view of $4.8 \arcmin \times 4.8 \arcmin$.
Image reduction and differential photometry were performed using a fully automated pipeline \citep{2015ASPC..496..181P} that incorporates Pyraf \citep{2012ascl.soft07011S} and the Astrometry.net packages \citep{2010AJ....139.1782L}.

J210228 was observed on five nights in total, three in 2019 (August 9 to 11 and 21) and two in 2021 (September 14 and 15). 2MASS~J21023075-0313205 (Johnson V-band magnitude V$_{\rm J}$ = 14.8 mag) was used as comparison star, 2MASS~J21023485-0309335 (V$_{\rm J}$ = 15.7 mag) as check star, and the photometric precision is estimated as 0.024-0.050 mag, 0.007-0.020 mag, 0.008-0.010 mag, and 0.009-0.020 mag in $BVRI$, respectively, depending on the observing conditions.

J231513 was observed on three nights in total (2019 September 21, 2020 October 11 and 2021 September 14). 2MASS~J23151130+3455434 (V$_{\rm J}$ = 15.5 mag) was used as comparison star, 2MASS~J23145819+3454092 (V$_{\rm J}$ = 15.5 mag) as check star, and the photometric precision is estimated as 0.008-0.012 mag, 0.006-0.008 mag and 0.007-0.010 mag in $VRI$, respectively, depending on the observing conditions.

\section{Light Curve modeling} \label{sect:modelling}
\begin{figure*}

\minipage{0.32\textwidth}
\includegraphics[width=\linewidth]{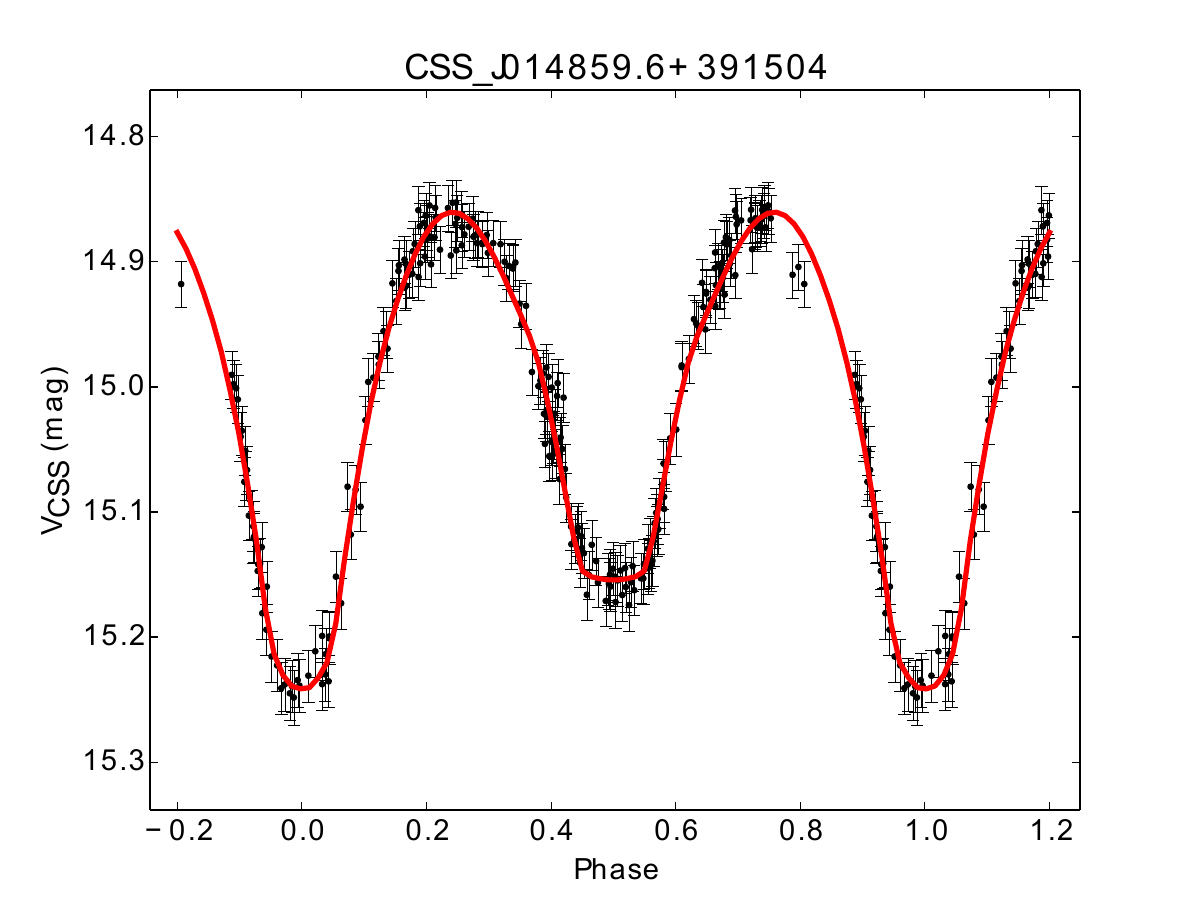} %
\endminipage\hfill
\minipage{0.32\textwidth}
\includegraphics[width=\linewidth]{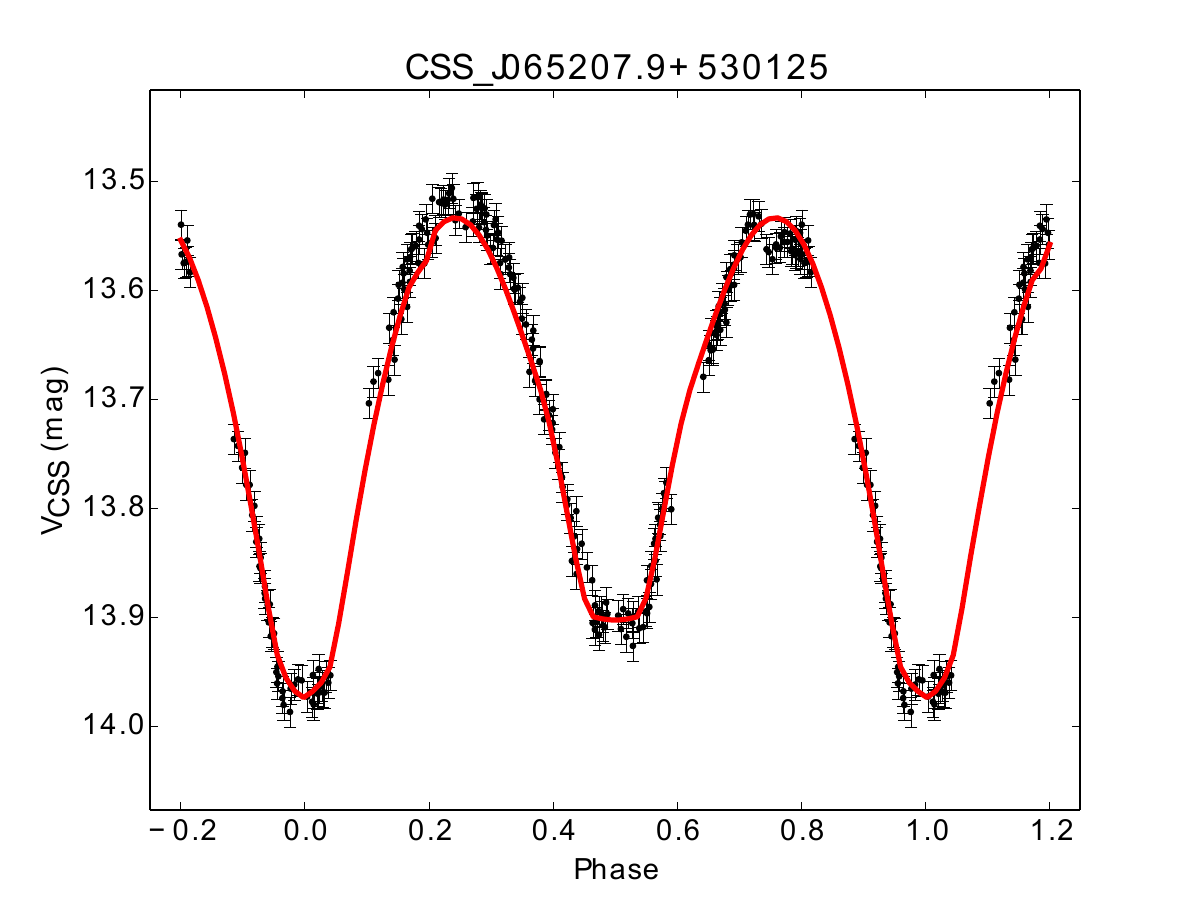} %
\endminipage\hfill
\minipage{0.32\textwidth}
\includegraphics[width=\linewidth]{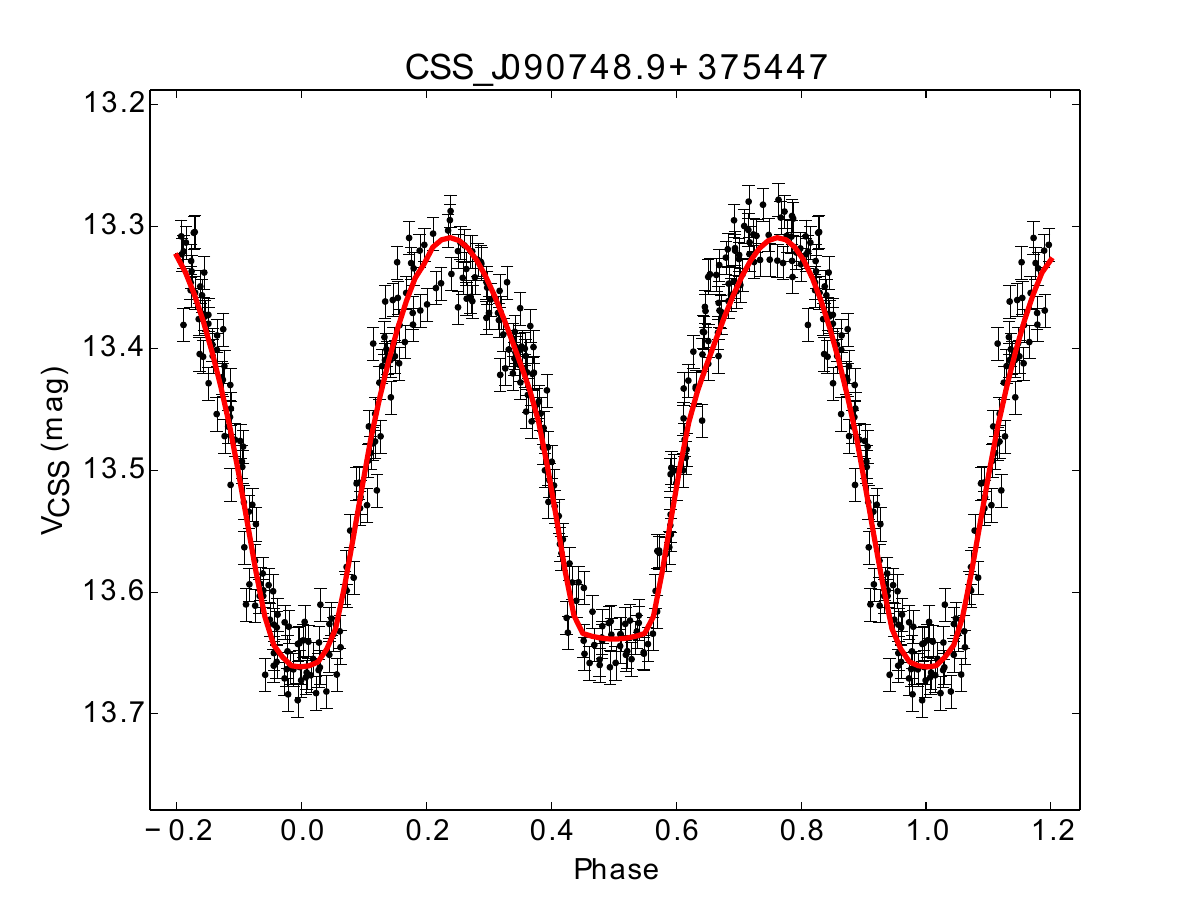} %
\endminipage
 
\minipage{0.32\textwidth}
\includegraphics[width=\linewidth]{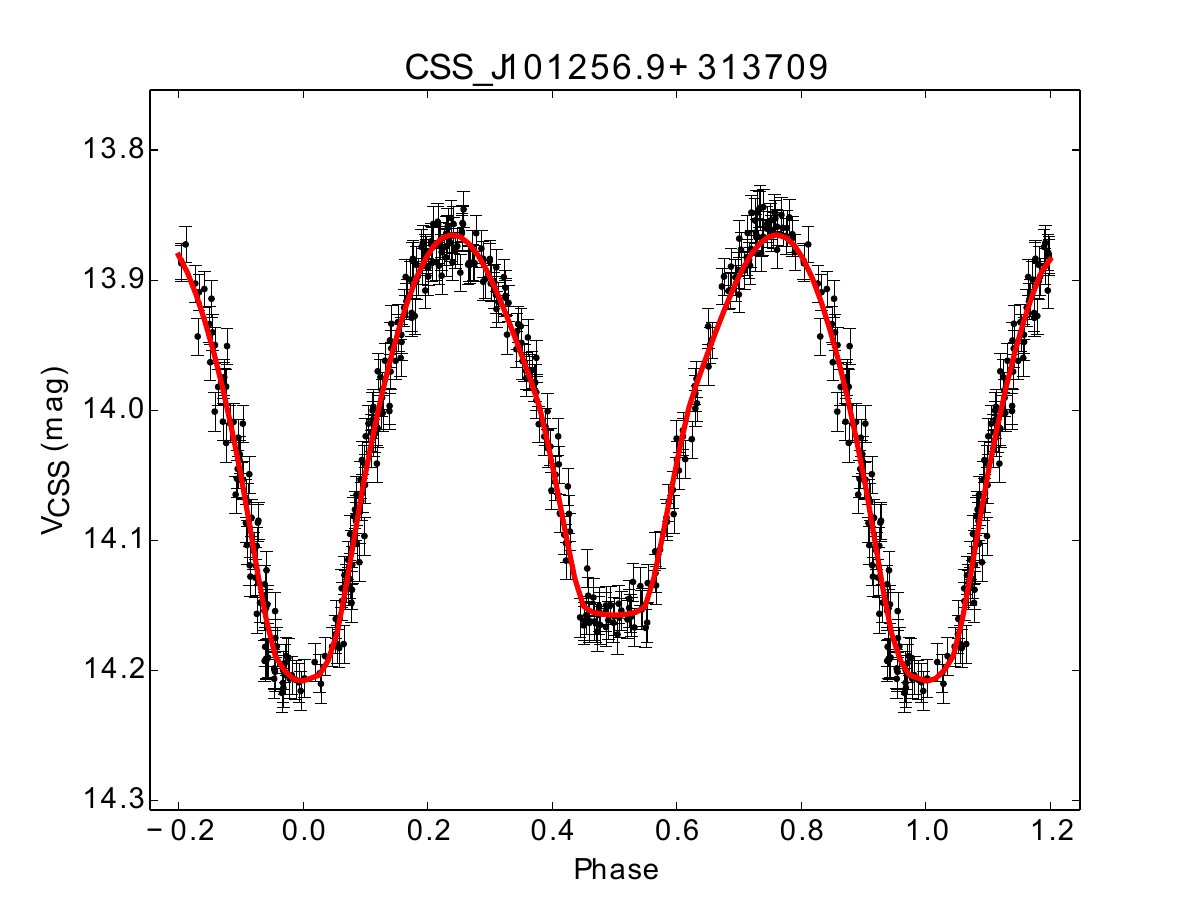} %
\endminipage\hfill
\minipage{0.32\textwidth}
\includegraphics[width=\linewidth]{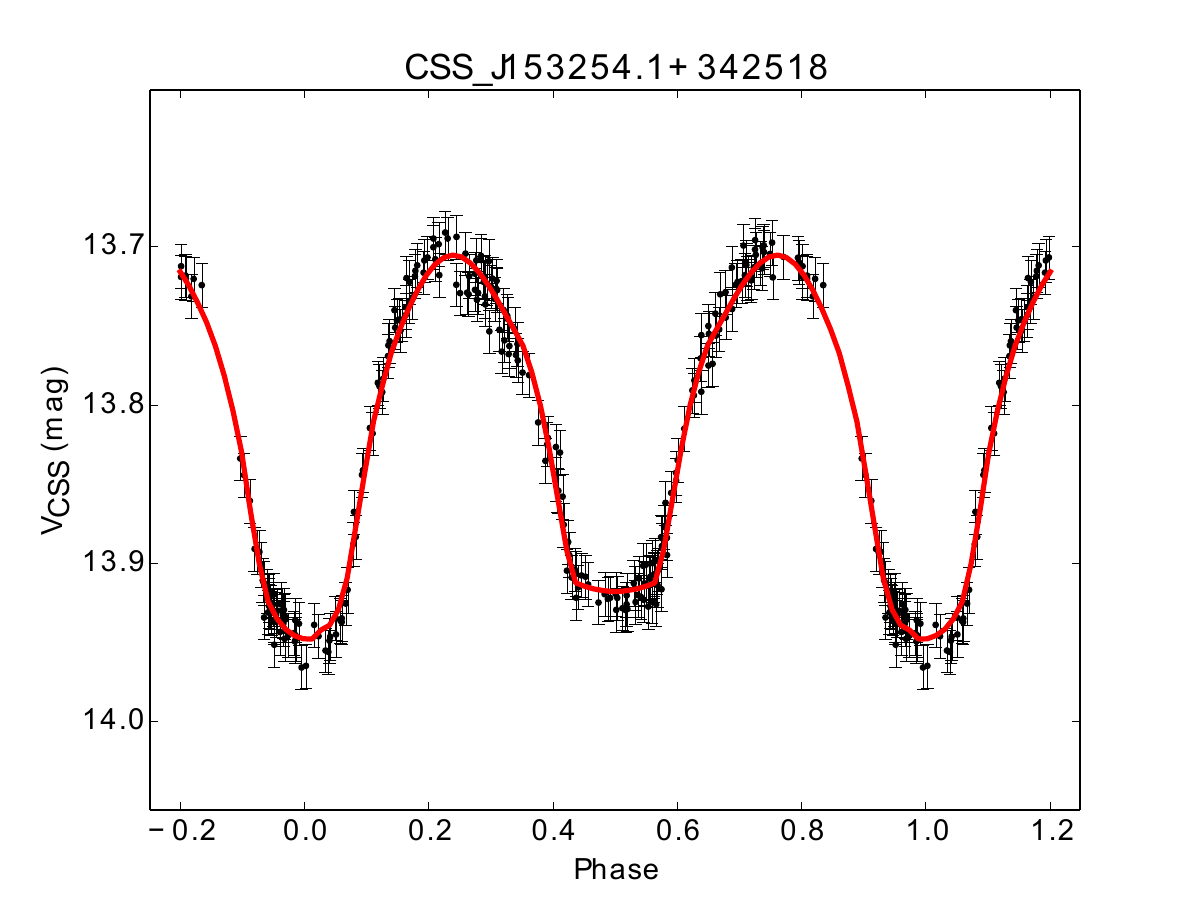} %
\endminipage\hfill
\minipage{0.32\textwidth}
\includegraphics[width=\linewidth]{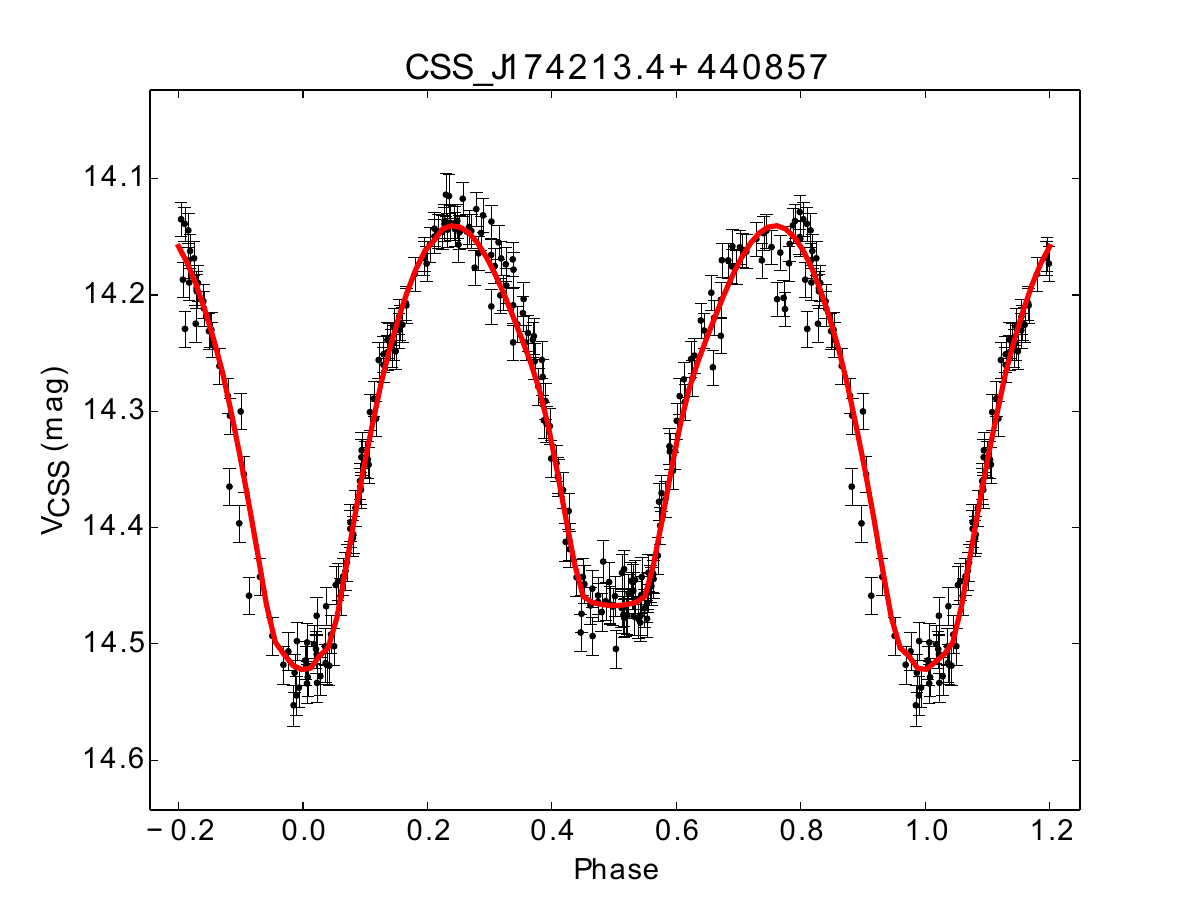} %
\endminipage
 
\minipage{0.32\textwidth}
\includegraphics[width=\linewidth]{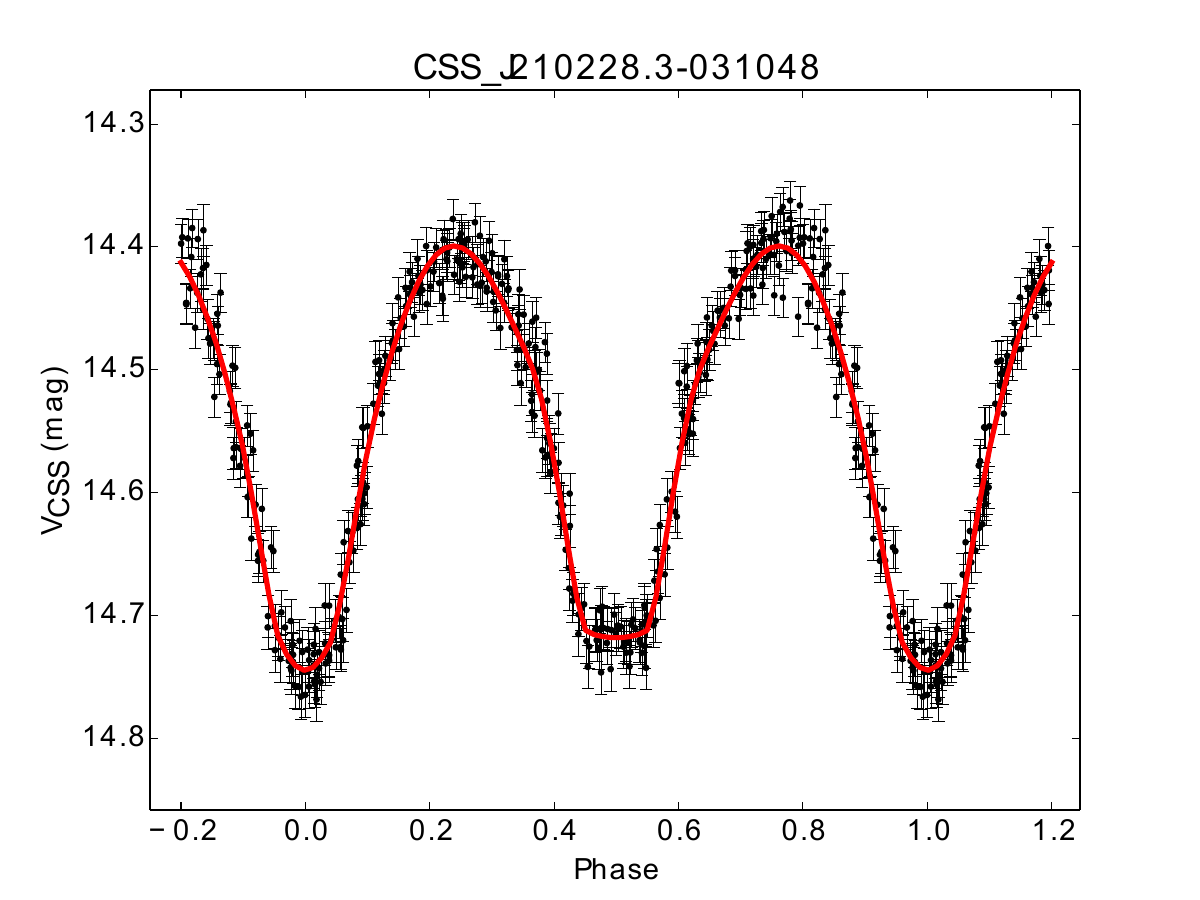} %
\endminipage\hfill
\minipage{0.32\textwidth}
\includegraphics[width=\linewidth]{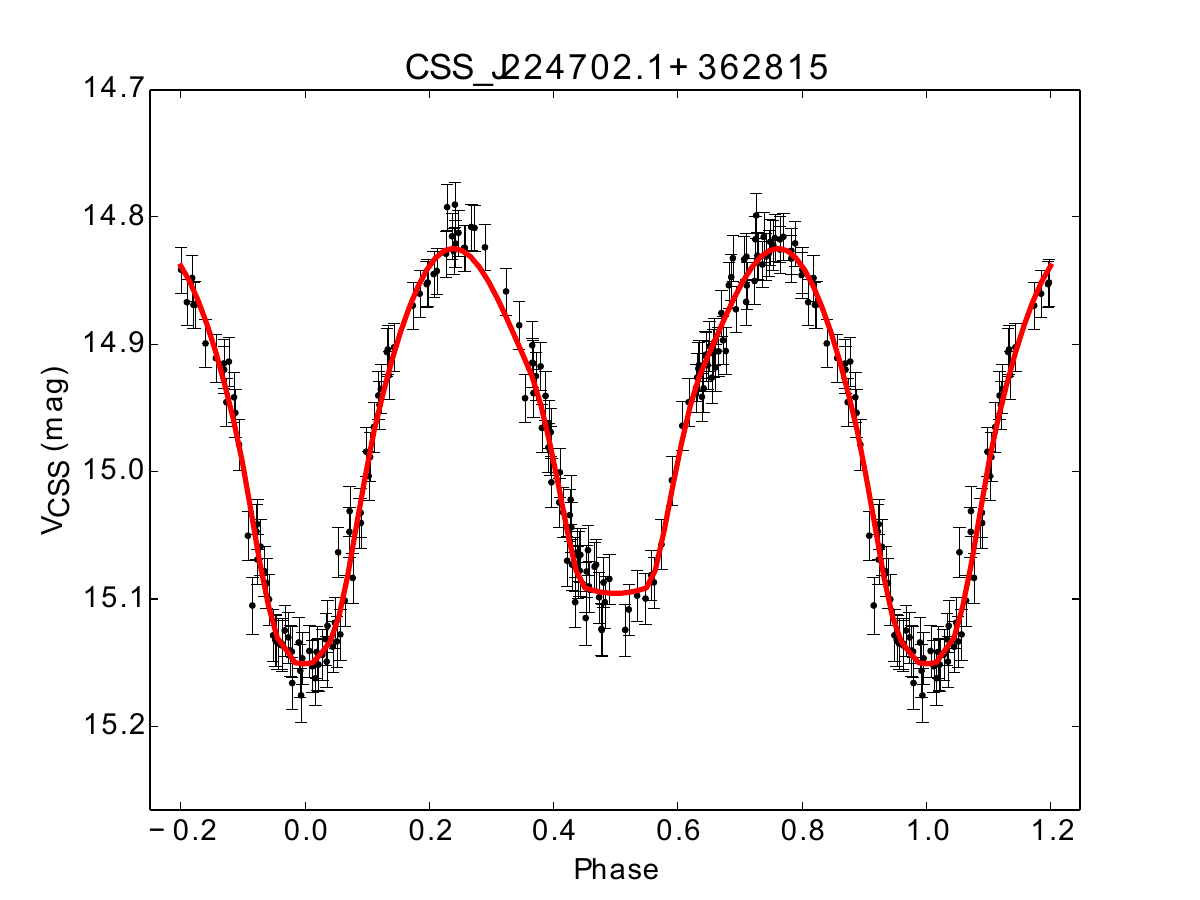} %
\endminipage\hfill
\minipage{0.32\textwidth}
\includegraphics[width=\linewidth]{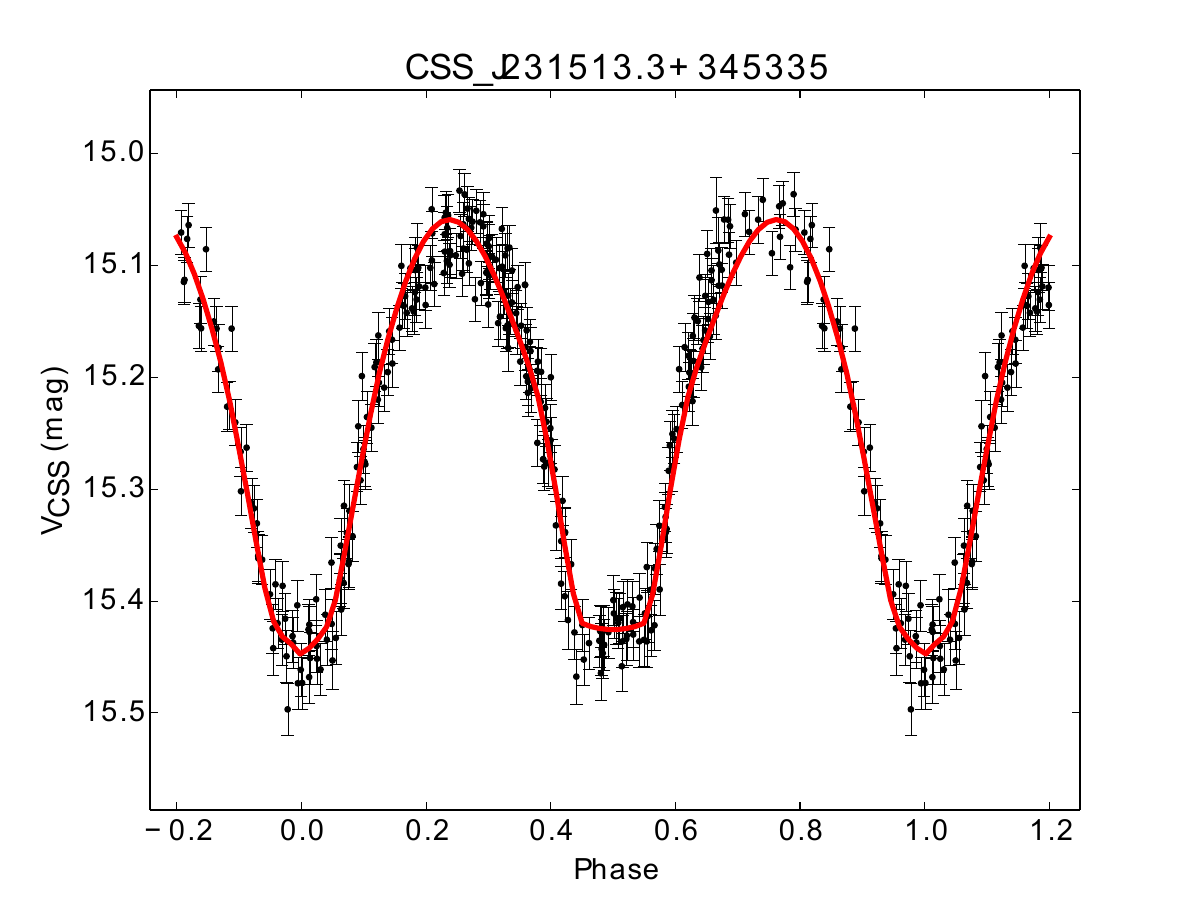} %
\endminipage
\caption{The observed LCs (black points) from CSS and the best-fitted models (red lines) for the 9 LMR systems.}\label{fig:LCsWsynt}
\end{figure*}

All 9 systems show total eclipses (Fig.~\ref{fig:LCsWsynt}). Totality, in the case of EWs, gives a reliable determination of the photometric mass ratio\footnote{Where $q=M_2/M_1$. In this work, we denote with index 1 the parameters of the more massive (primary) and 2 those of the less massive (secondary) component.}, $q$ \citep{Terrell&Wilson2005,Hambalek2013,Senavci2016} when radial velocities are not available. Thus, the CSS and the new multiband LCs were analysed using the PHOEBE-0.31a scripter \citep{2005ApJ...628..426P} in ``Overcontact not in thermal contact'' mode. Since the LC morphology constrains only the temperature ratio ($T_{2}/T_{1}$) of the two components, an estimation of the primary's effective temperature $T_1$ is needed. We adopt the effective temperatures of the systems ($T_{\rm sys}$) from \cite{2019AJ....158..138S}, and assuming that the $T_{\rm sys}$ is dominated by the temperature of the hotter component ($T_{1}$), we set $T_{1} = T_{\rm sys}$ as a fixed value. %This was derived from either spectroscopy or from empirical relations between $T_{\rm eff}$ and the {\em Gaia} $G_{\rm BP} - G_{\rm RP}$ color \citep[DR2;][]{GAIADR2}, based on a set of 19962 stars having spectroscopically determined $T_{\rm eff}$ and being within 100 pc to avoid reddening.
For  J065207, J090748, and J101256 the adopted spectroscopic value of effective temperature is in agreement with the Large Sky Area Multi-Object Fiber Spectroscopic Telescope low-resolution spectrum \citep[LAMOST DR7;] []{2012RAA....12.1197C}   data \citep{2020RAA....20..163Q}, and for J174213 from LAMOST observations (01-05-2017) within errors. For  J014859, J210228, J224702, and J231513 there are no spectroscopic data whereas for  J153254 we set the effective temperature 6747$\pm$45 K from LAMOST (9-3-2015). The limb darkening coefficients were interpolated from \cite{1993AJ....106.2096V} tables for the given temperatures, while the bolometric albedos and the gravity darkening coefficients were set as $A_1=A_2=$ 0.5 \citep{1973AcA....23...79R} and $g_1=g_2=$ 0.32 \citep{1967ZA.....65...89L}, respectively, for $T_{\rm sys}$ < 7200~K, and $A_1=A_2=$ 1 and $g_1=g_2=$ 1 otherwise. The synchronicity parameter was set as $F_{1,2}=1$, which corresponds to a synchronous rotation, in a circular orbit. Finally, the ephemeris of each system was improved by adjusting the period and the reference time of minimum, $\rm HJD_0$ (see Table~\ref{tab:1}).

Our next step was to perform a 2D grid search on the mass ratio - inclination ($q-i$) plane in order to determine the pair ($q_{\rm min},i_{\rm min}$) that minimizes the $\chi^2$ of the observed minus calculated LC. In a first attempt to estimate ($q_{\rm min},i_{\rm min}$), the range of $q$ was set at [0.1-0.6], with an increment of 0.05, and the range of $i$ at [65$^\circ$-90$^\circ$], with an increment of 1$^\circ$. This was executed into two runs, one with phase shift 0.0 and a second with 0.5 to investigate values of $q\leq1$ and $q>1$, respectively.
Then, depending on the first results, the explored $q$ values were set in the range of $q_{\rm min}\pm 0.05$, with an increment of 0.01.
The derived physical parameters from the fitting, namely mass ratio $q$, temperature ratio $T_2/T_1$, modified equipotential $\Omega_{1,2}$, relative radii of the components $r_{1,2}$, inclination $i$, bandpass luminosity of the primary component over the total bandpass luminosity $L_{1}/L_{\rm tot}$, and fillout factor\footnote{$f=\frac{\Omega-\Omega_{\rm in}}{\Omega_{\rm out}-\Omega_{\rm in}}$, where $\Omega_{\rm in}$ and $\Omega_{\rm out}$ are the modified potential at the inner ($L_1$) and the outer ($L_2$) Lagrangian points, respectively.} $f$, are listed in Table~\ref{tab:models}. The third light contribution was also investigated and found to be negligible ($\leq 1\%$) given the photometric precision of CSS LCs.

\begin{figure*}
\minipage{0.5\textwidth}
\centering
\includegraphics[width=\linewidth]{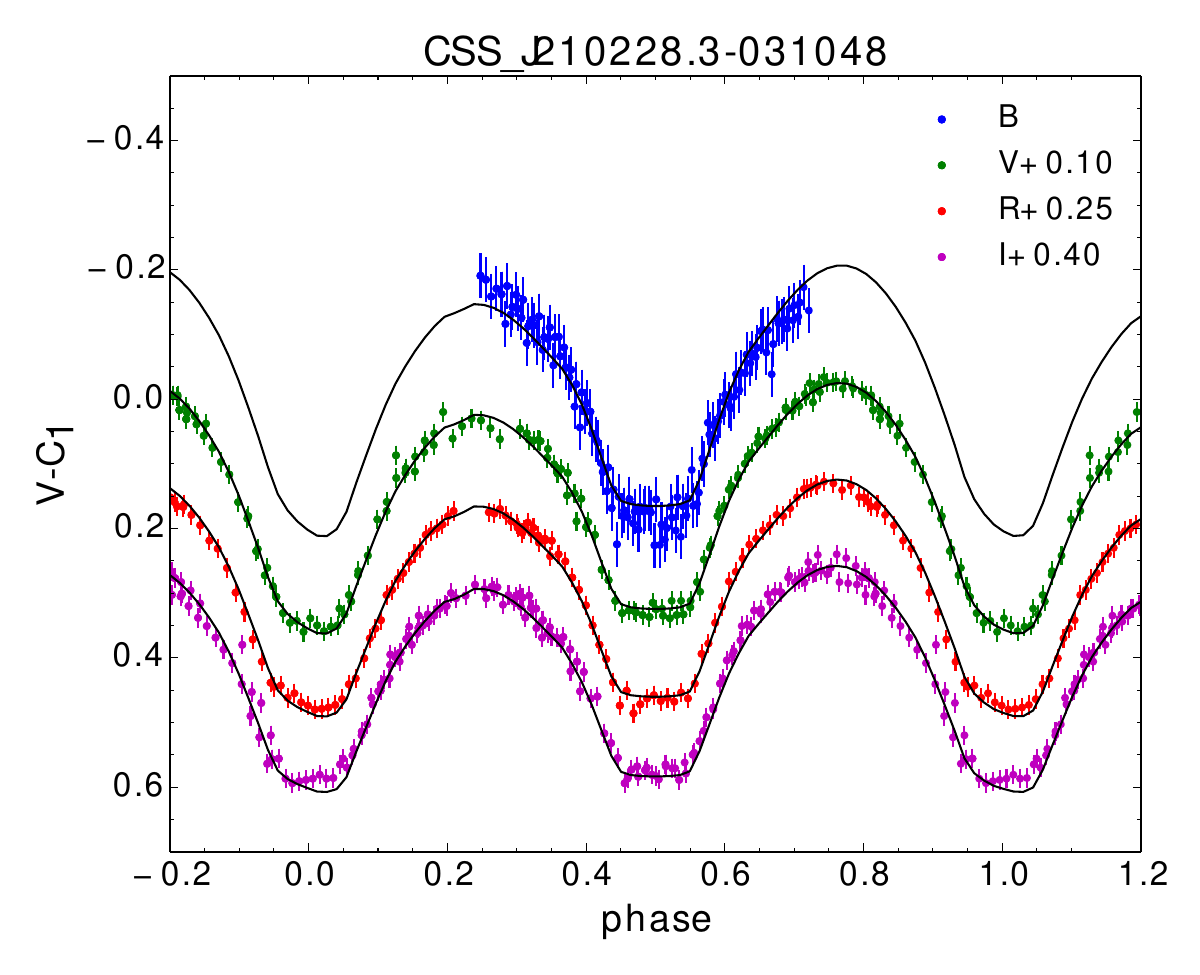} %

\endminipage\hfill
\minipage{0.5\textwidth}
\includegraphics[width=\linewidth]{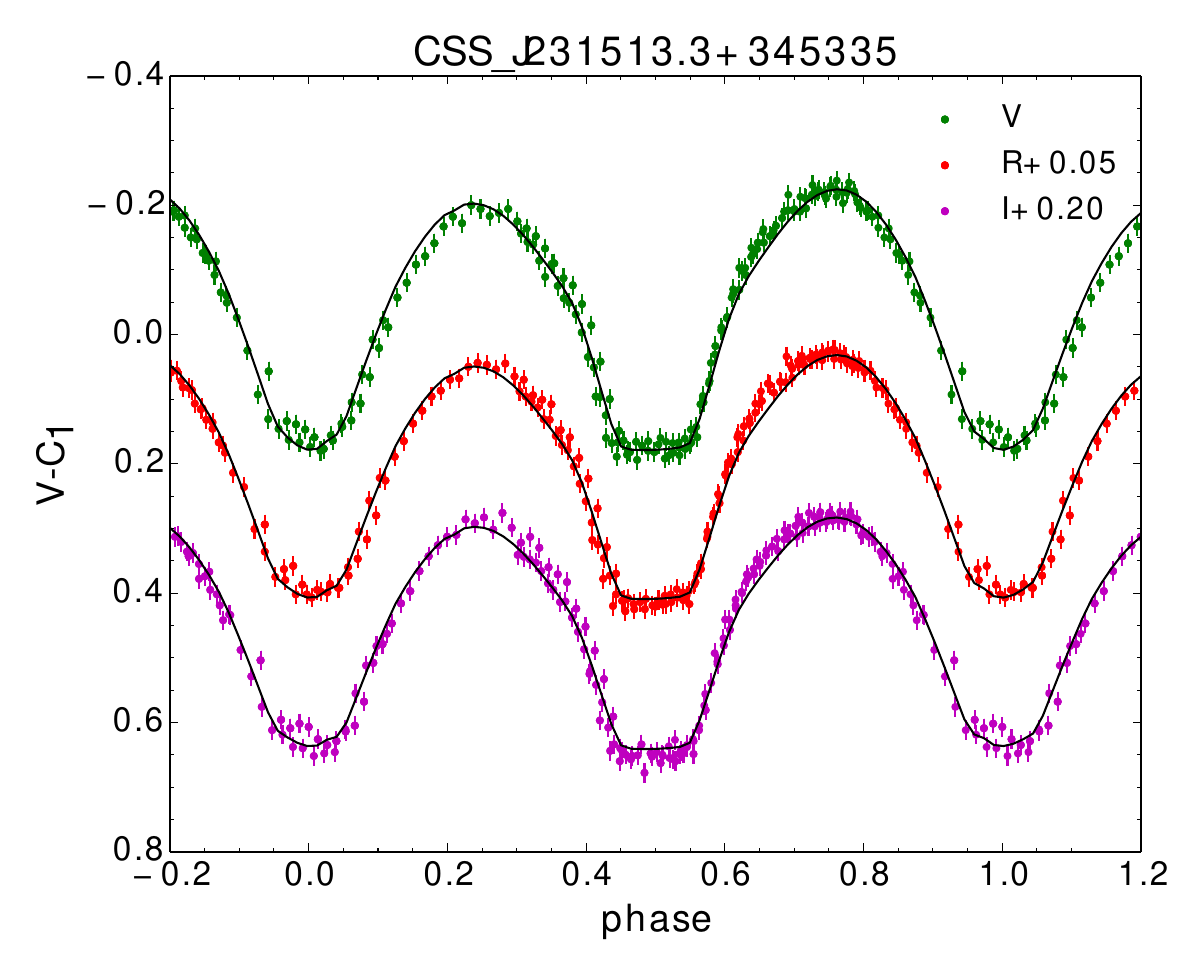} %
\endminipage
\vspace{+0.3cm}
\caption{Phase-folded $BVRI$ (left) and $VRI$ (right) observations for J210228 and J231513, respectively, with their models (black line).}\label{fig:BVRILC}
\end{figure*}

We chose to model the new multiband LCs of J210228 and J231513 independently from that of CSS to investigate whether the two solutions agree. The results are also presented in Table~\ref{tab:models} and show that the modified equipotential $\Omega_{1,2}$ and subsequently the fillout factor $f$, as well as the temperature ratio $T_2/T_1$, are mainly affected, while the mass ratio $q$ and the inclination $i$ is the same within the errors (see Sect.~\ref{subsect:errors}). This was also pointed out by \cite{2023MNRAS.519.5760L,2023AN....34420066W, 2024MNRAS.527.6406L}.
Fig.~\ref{fig:BVRILC} presents the observed LCs and the best-fitted model (black line). As one can clearly see from Fig.~\ref{fig:BVRILC}, the dedicated multiband observations reveal an asymmetry at the maxima of both J210228 and J231513, suggestive of the O'Connell effect \citep{1951MNRAS.111..642O} possibly caused by chromospheric spots on the surface of the components or one hot spot due to the mass transfer. 
In the case of J210228, two spots were needed to describe the observed LCs: a hot one on the primary component (co-latitude 100$\degr$ and longitude 15$\degr$) and a cool one on the secondary (co-latitude 100$\degr$ and longitude 15$\degr$). The hot spot has a radius of 20$\degr$ and a temperature factor of $T_{\rm s}/T_1=1.05$ (where $T_{\rm s}$ is the temperature on the spot surface) while the cool spot has a radius of 30$\degr$ and a temperature factor of $T_{\rm s}/T_2=0.75$.
   
In the case of J231513, one cool spot ($T_{\rm s}/T_1=0.96$), with a radius of 20$\degr$ on the primary (co-latitude 90$\degr$ and longitude 270$\degr$) can satisfactorily describe the observed LCs.

\subsection{Error estimation of the physical parameters}\label{subsect:errors}
The errors provided in Table~\ref{tab:models} were estimated using Monte-Carlo simulations \citep{2015AJ....149..168P,PA19}. During this procedure, 1000 synthetic LCs are produced from the observed ones by a random displacement of each photometric point. The displacement depends on the photometric error ($\sigma _{\rm CSS}$) of the corresponding point (a value selected randomly from a normal distribution with zero mean and $\sigma_{\rm CSS}$ as standard deviation). Finally, the errors are extracted from each parameter's ($T_2/T_1, \Omega_{1,2}, r_1, r_2, i$) distribution within 3~$\sigma$.

\begin{figure*}
\minipage{0.4\textwidth}
\centering
\includegraphics[width=\linewidth]{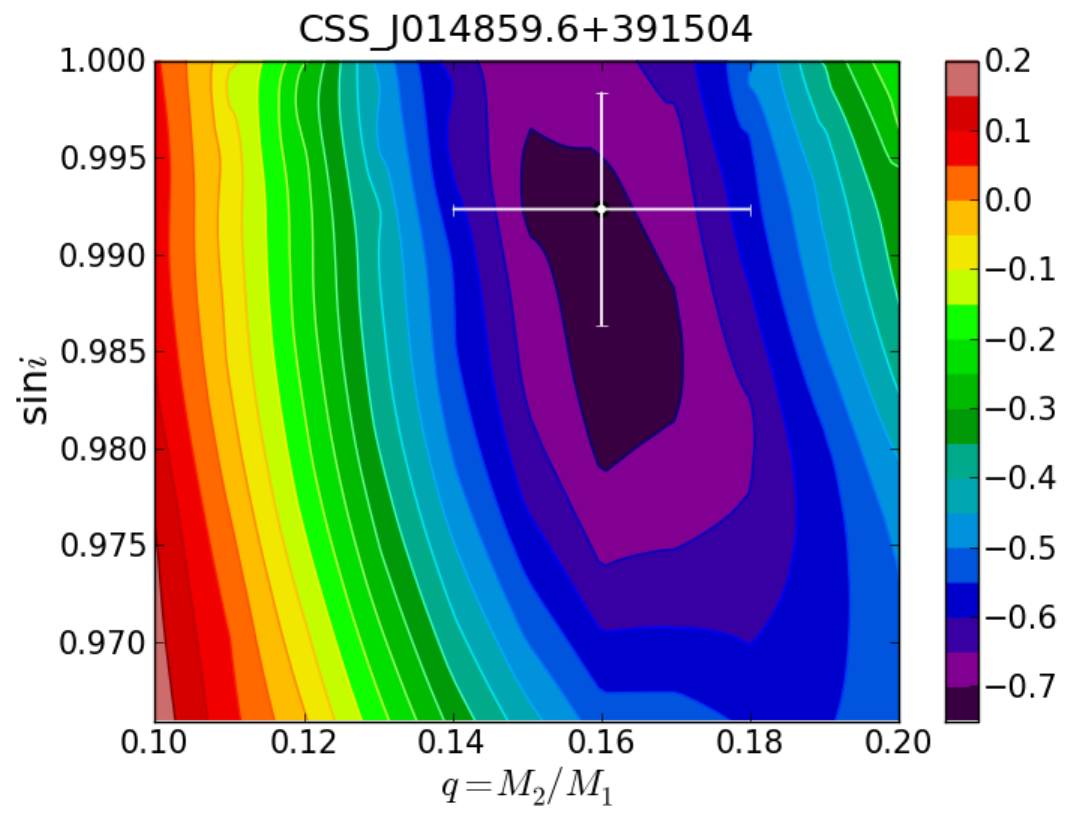} %
\endminipage%\hfill
\hspace{+1.5cm}
\minipage{0.4\textwidth}
\includegraphics[width=\linewidth]{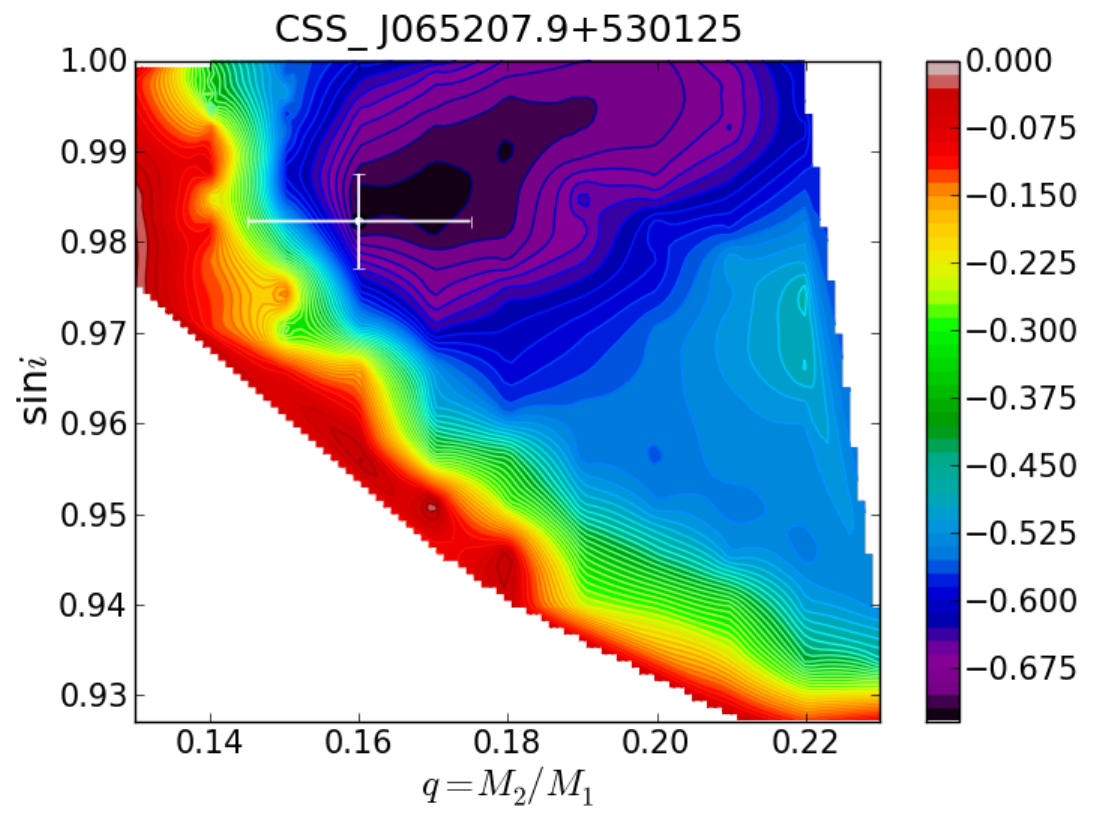} %
\endminipage
\vspace{+0.3cm}

\minipage{0.4\textwidth}
\centering
\includegraphics[width=\linewidth]{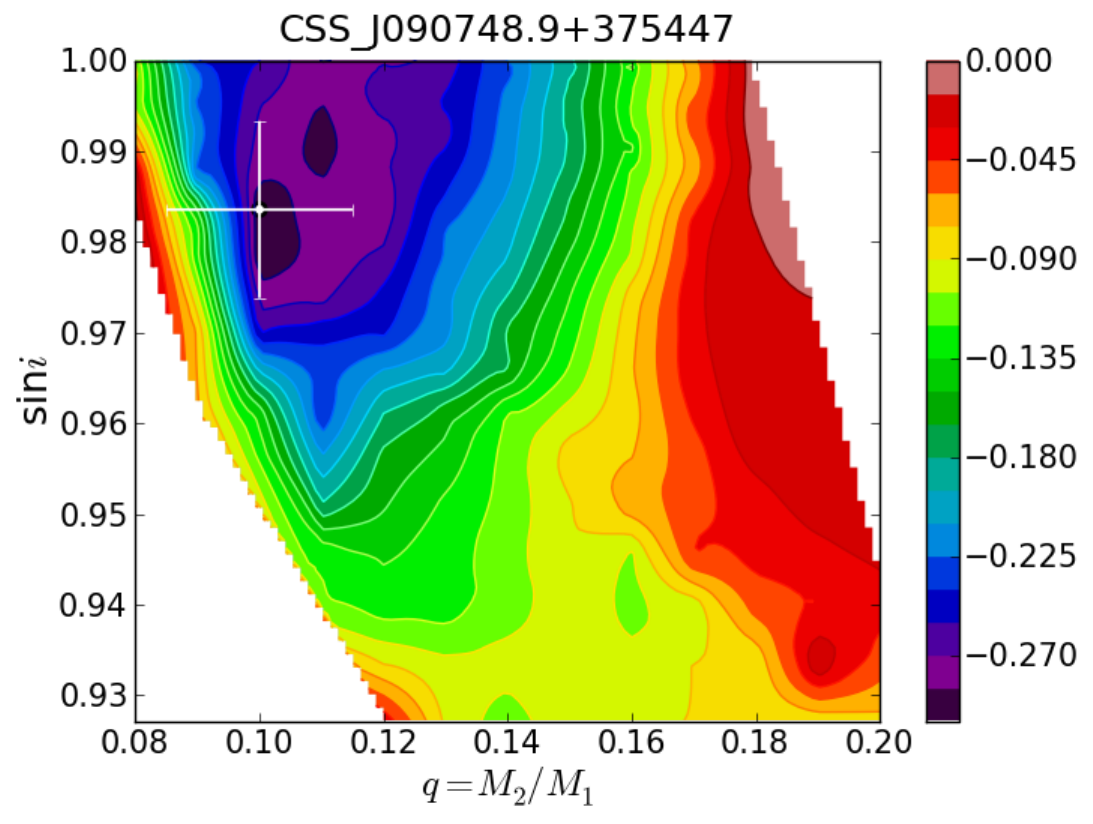} %
\endminipage%\hfill
\hspace{+1.5cm}
\minipage{0.4\textwidth}
\includegraphics[width=\linewidth]{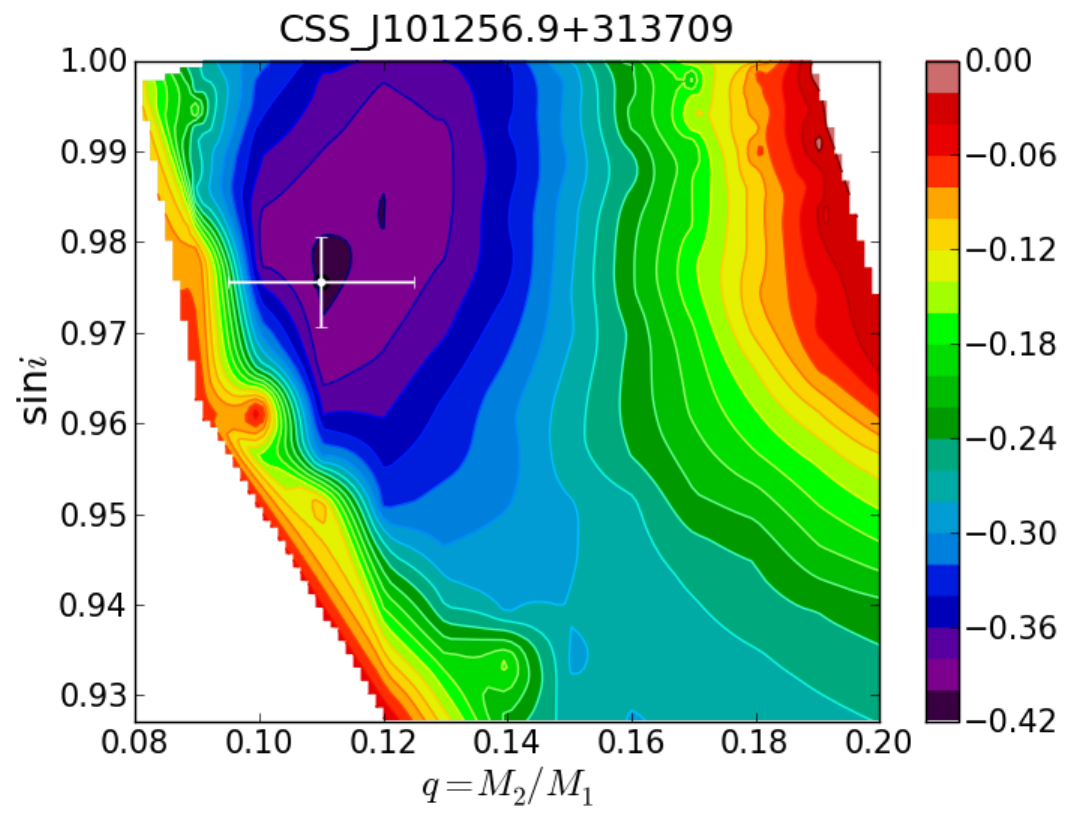} %
\endminipage
\vspace{+0.3cm}

\minipage{0.4\textwidth}
\centering
\includegraphics[width=\linewidth]{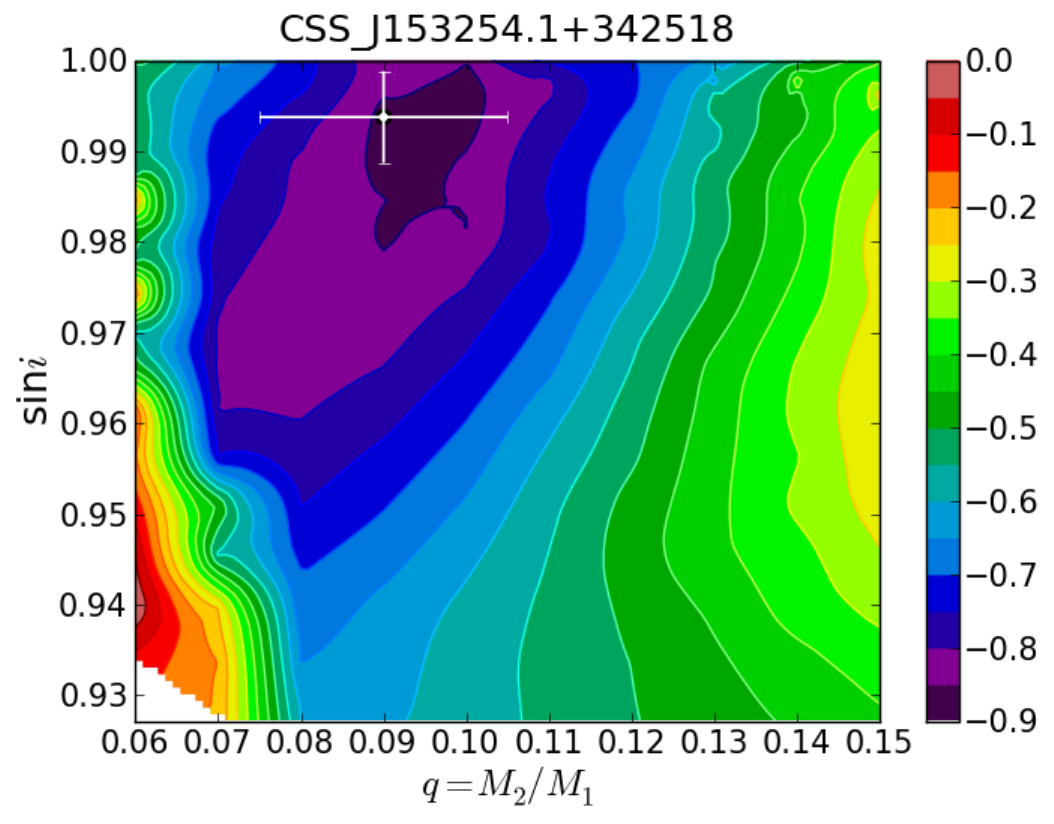} %
\endminipage%\hfill
\hspace{+1.5cm}
\minipage{0.4\textwidth}
\includegraphics[width=\linewidth]{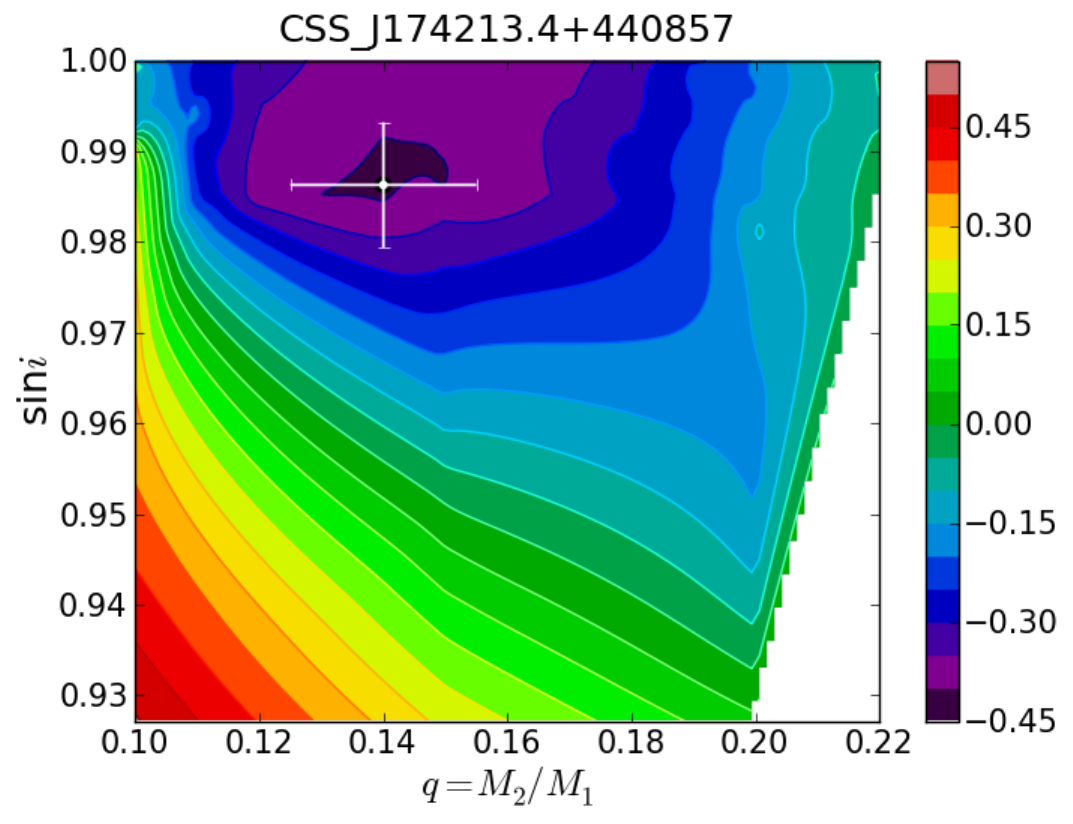} %
\endminipage
\vspace{+0.3cm}

\minipage{0.4\textwidth}
\resizebox*{\hsize}{!}
            {\includegraphics[width=\textwidth]{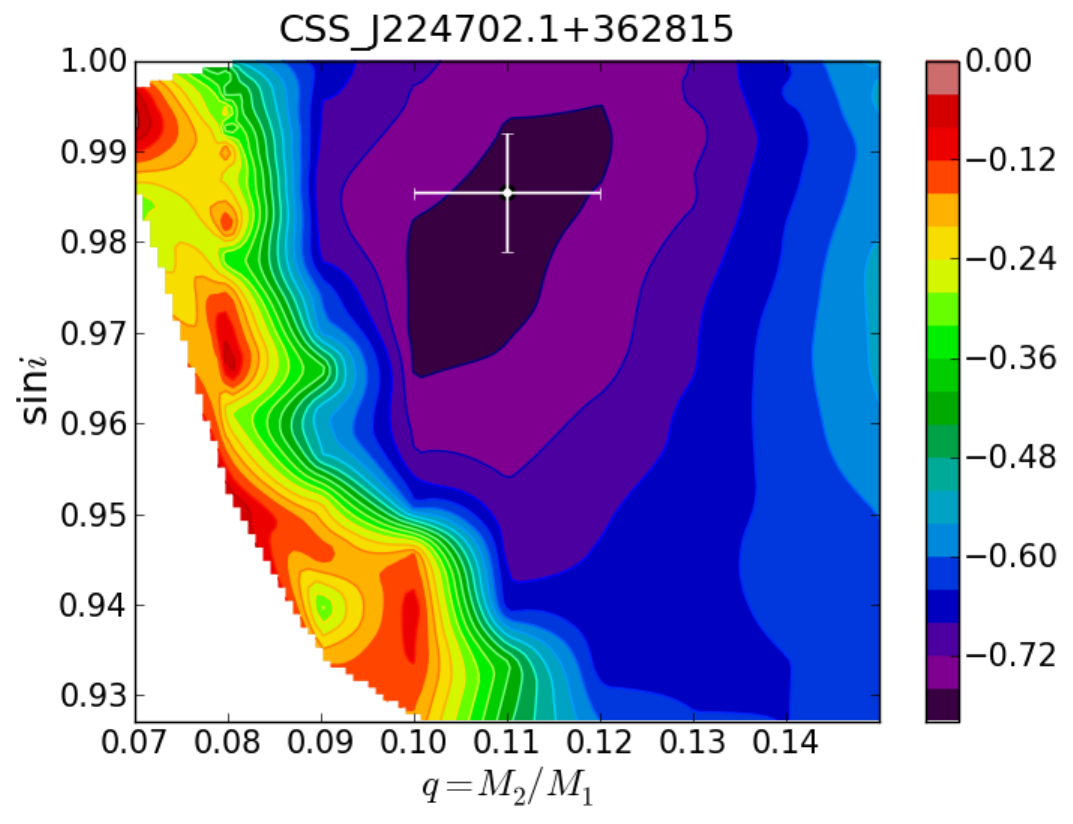}}
\endminipage
\vspace{+0.3cm}
\caption{Contour plots of $\log{\chi^2}$ (color-coded according to the scale on the right) on the ($q$, $\sin i$) plane as resulted from the $q - i$ scan using the CSS LC. The white crosses represent the solutions of the systems with the error bars derived from the MC and heuristic scanning, respectively. The size of the error bars is drawn according to the errors.}\label{fig:contours}
\end{figure*}

\begin{figure*}
\minipage{0.4\textwidth}
\centering
\includegraphics[width=\linewidth]{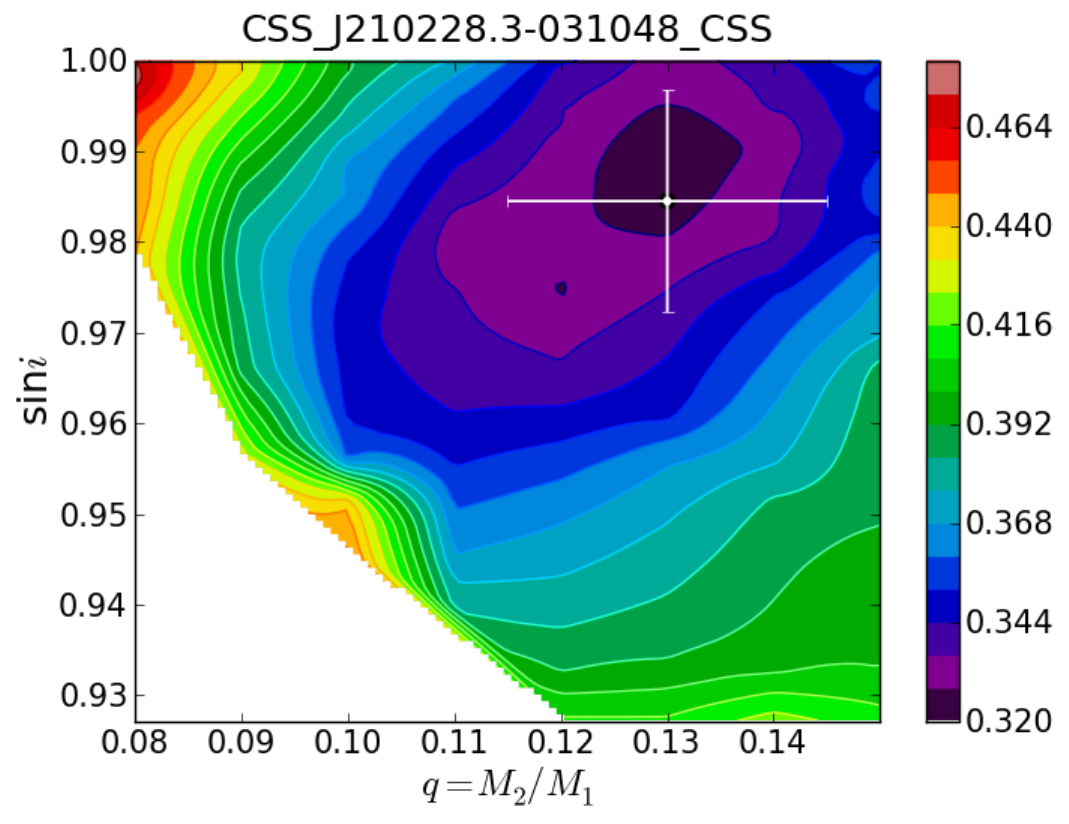} %
\center \textbf{(a)}
\endminipage
\minipage{0.4\textwidth}
\hspace{+1.5cm}
\includegraphics[width=\linewidth]{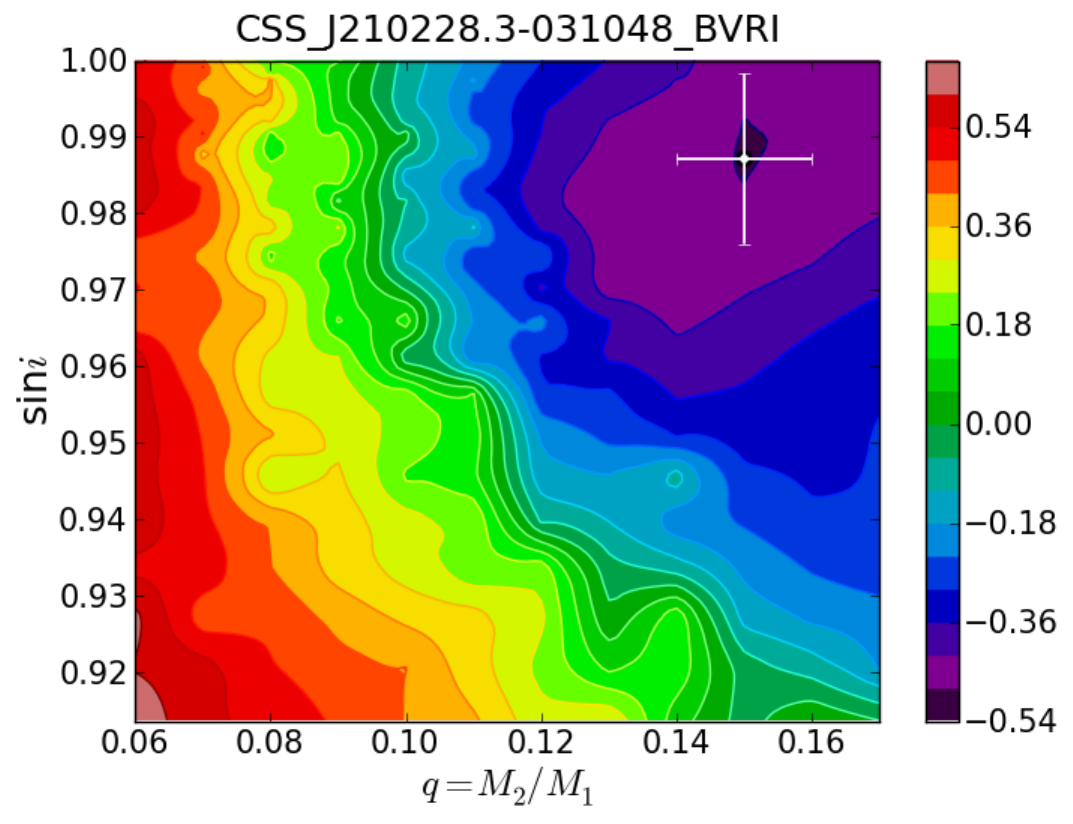} %
\center\hspace{+2.8cm} \textbf{(b)}
\endminipage
\vspace{+0.3cm}
\minipage{0.4\textwidth}
\centering
\includegraphics[width=\linewidth]{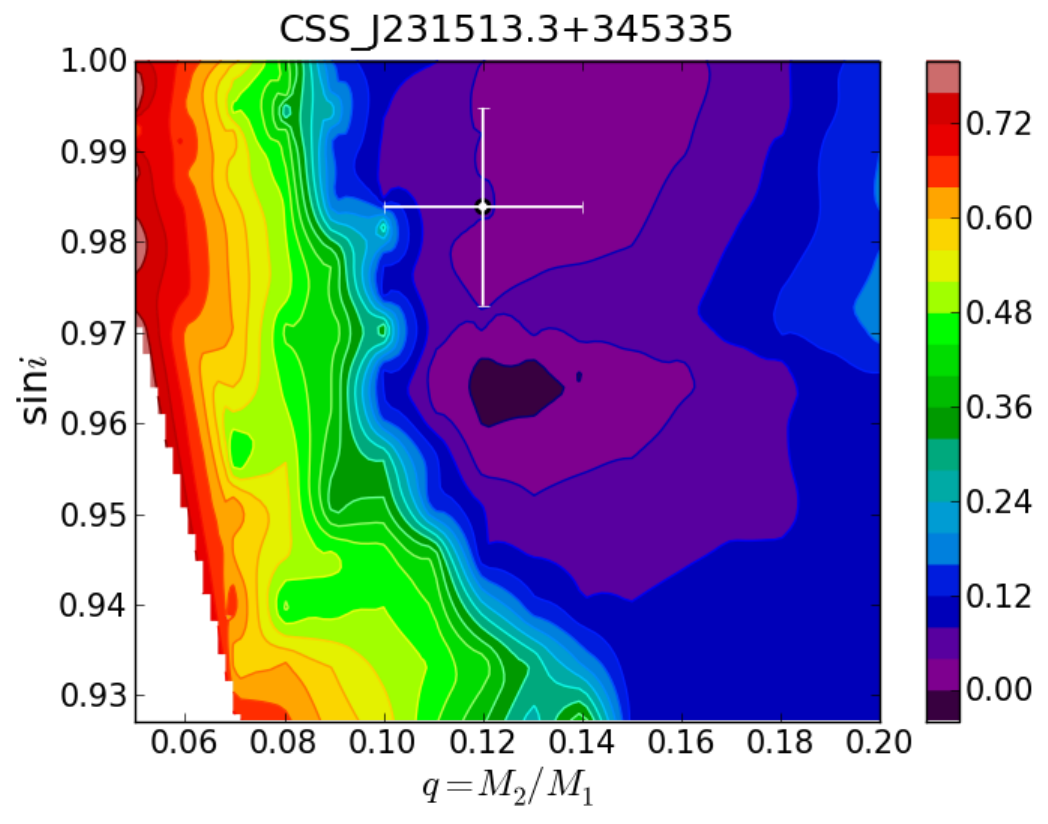} %
\center \textbf{(c)}
\endminipage%\hfill
\minipage{0.4\textwidth}
\hspace{+1.5cm}
\includegraphics[width=\linewidth]{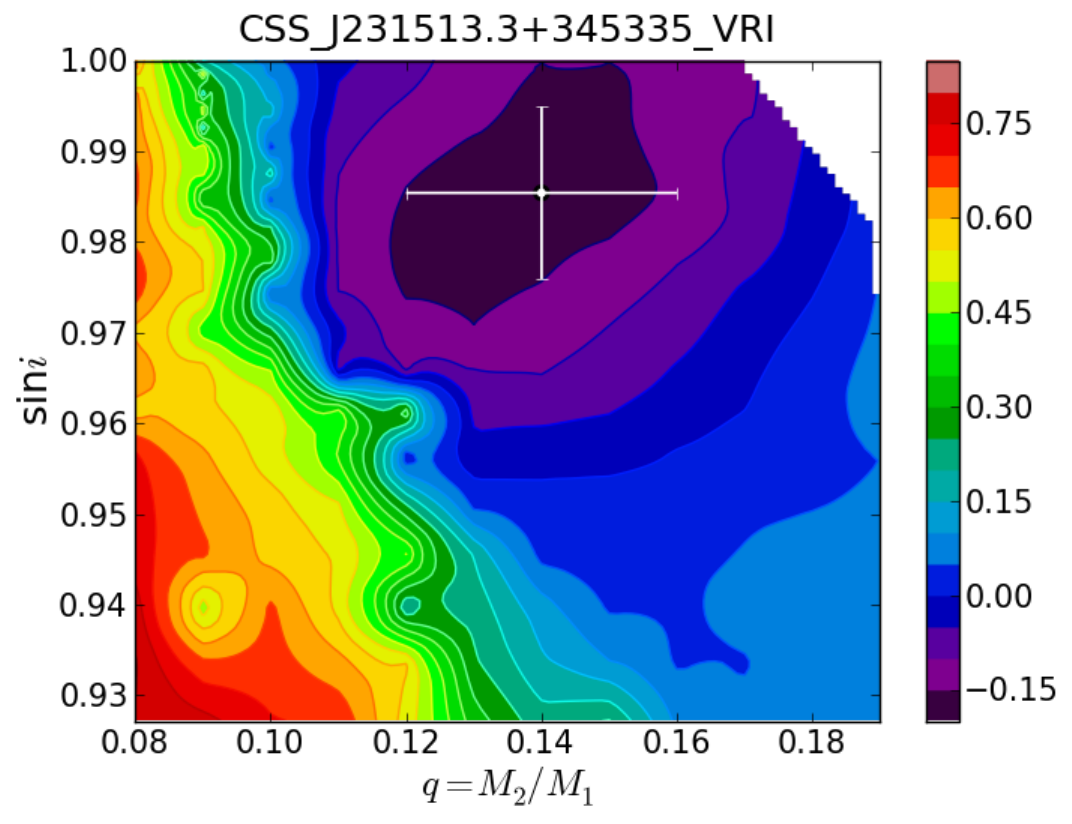} %
\center\hspace{+2.8cm} \textbf{(d)}
\endminipage
\vspace{+0.3cm}
\caption{Same as Fig.~\ref{fig:contours} but for J210228 (upper panel) and J231513 (lower panel) using CSS (left) and new $BVRI$ observations (right).}\label{fig:qi_contours_obj2}
\end{figure*}

In the case of the new multiband observations of J210228 and J231513, the errors were estimated using parameter kicking \citep{2005ApJ...628..426P,PapageorgiouPhD} with the PHOEBE-0.31a scripter. This procedure achieves the same effect as stochastic methods (e.g., simulated annealing) but in a significantly shorter time \citep{10.1088/978-0-7503-1287-5}. Starting from the solution derived by the $q-i$ scan, the parameters $T_2/T_1, \Omega_{1,2}, i, L_{1}$ are adjusted (80 iterations) and then perturbed by a factor of $5\%$ of their value. The updated model is used as the initial one in order to start a second run, and the loop is repeated 100 times. The errors of the parameters ($T_2/T_1, \Omega_{1,2}, r_1, r_2, i$) are estimated from their distribution within 3~$\sigma$. In this way, it is possible to escape the local minima and test the stability of the solution. Nevertheless, the errors of some physical parameters, particularly the fillout factor f, are underestimated. This may result, as mentioned earlier, from the difference in photometric accuracy between survey and dedicated data or among different surveys.
The mass ratio uncertainties $\delta q$ were estimated from the $q-i$ scan following the method described by \cite{2022MNRAS.512.1244C}.
Figs.~\ref{fig:contours} and~\ref{fig:qi_contours_obj2} present the log $\chi^2$ topology on the $q- \sin i$ plane derived from the $q-i$ scan. The white cross shows the solution along with its errors derived from heuristic scanning.

\subsection{Absolute parameters estimation}
Having determined the temperature ($T_2/T_1$) and the radii ($r_2/r_1$) ratio of the systems' components, we use the relation of \cite{2003A&A...404..333Z} to derive the individual effective temperatures $T_1$, $T_2$ (Table~\ref{tab:absparams}). Then, in order to estimate the absolute parameters of the 9 systems, we use our semi-empirical mass-luminosity relation for the primary component \citep{2022MNRAS.512.1244C, 2023AJ....165...80P}, namely:
\begin{equation}
 \label{eq:1}
        \log L_{1}=\log (0.63 \pm 0.04)+(4.8 \pm 0.2)\log M_{1}.   
\end{equation}
This relation is derived using a sample of 161 spectroscopically studied EWs with well-determined absolute parameters, by performing a linear fitting on the $\log M_1-\log L_1$ plane \citep[see][]{2023AJ....165...80P}. We estimate the luminosity of the primary component $L_1$ using the magnitude at the maxima of the LCs (Table~\ref{tab:1}) and the luminosity ratio $L_1/L_{\rm tot}$ (Table~\ref{tab:models}) as derived from the modeling, taking advantage of the new distances from {\em Gaia} Data Release 3 \citep[DR3;][]{2023A&A...674A...1G}. All the 9 systems have well-determined distances with uncertainties $\lesssim$ 6\% of their value, except J014859 (15$\%$).  
The CSS magnitudes were transformed to $V$ Johnson according to the relation from \cite{2013ApJ...763...32D} using B-V color indices from $\it{TESS}$ Input Catalog, v-8.0 \citep[][]{2019AJ....158..138S}.
The color extinction $A_V$ was adopted from \textsc{mwdust} \citep{2016ApJ...818..130B,2019ApJ...887...93G} for the given distances, while the required bolometric corrections are obtained from \citet[][version 2022.04.16]{2013ApJS..208....9P} for the given temperature of the primary. Following the determination of the mass of the primary ($M_1$), the mass of the secondary $M_2$ is calculated using the mass ratio $q$, while the radii ($R_1$, $R_2$) of the components are derived using Kepler's law and the relative radii obtained from modeling (Table~\ref{tab:models}). The results of this procedure are summarized in Table~\ref{tab:absparams}, along with the errors in the derived parameters, derived using error propagation.
We use the solutions obtained from our data for the absolute parameters estimation of J210228 and J231513.
\begin{table*}
\centering
\caption{Basic information for the 9 LMR EW systems from CSS.}
\label{tab:1}
\hspace{-1.5cm}
\begin{tabular}{cccccccccc}\hline
ID & Short ID & ${\rm RA}_{J2000}$  &  ${\rm Dec}_{J2000}$  &  $\rm HJD_0$  & Period & $V_{\rm CSS}({\rm max})$ & $\sigma_{\rm CSS}$ & $N_{\rm p}$    \\ 
& &(h:m:s) & (\degr:\arcmin:\arcsec)  &  (${\rm 2450000+}$) & (days) & (mag) & (mag) & \\
 \hline

   CSS$\_$J014859.6+391504 & J014859 & 01:48:59.65 & +39:15:04.2 & 5479.72138     & 0.7777702 & 14.86 & 0.02   & 271 \\
   CSS$\_$J065207.9+530125 & J065207&06:52:07.92 & +53:01:25.6 & 5093.00000     & 0.7608190 & 13.53 & 0.01   & 293 \\
   CSS$\_$J090748.9+375447 & J090748 & 09:07:48.94 & +37:54:47.3 & 6404.69949     & 0.3092479 & 13.31 & 0.01   & 380 \\
   CSS$\_$J101256.9+313709 & J101256 &10:12:56.97 & +31:37:09.6 & 4096.81447     & 0.3627181 & 13.86 & 0.01   & 408 \\
   CSS$\_$J153254.1+342518 & J153254 & 15:32:54.17 & +34:25:18.1 & 4157.85294     & 0.4550895 & 13.71 & 0.01   & 286 \\
   CSS$\_$J174213.4+440857 & J174213 & 17:42:13.49 & +44:08:57.3 & 3507.86169     & 0.4245780 & 14.14 & 0.02   & 286 \\
   CSS$\_$J210228.3-031048 & J210228 & 21:02:28.30 & -03:10:46.9 & 6208.99231     & 0.2454814 & 14.40 & 0.02   & 421 \\
   CSS$\_$J224702.1+362815 & J224702 &22:47:02.10 & +36:28:15.8 & 4953.96243     & 0.4849898 & 14.82 & 0.02   & 212 \\
   CSS$\_$J231513.3+345335 &J231513 &23:15:13.35 & +34:53:35.9 & 5985.73362     & 0.2661317 & 15.06 & 0.02   & 315 \\
\hline
\end{tabular}
\end{table*}
%%%%%%%%%%%%%%%%%%%%%%%%%%%%%%%%%%%%%%%%%%%%%%%%%%%%%%%%%%%%%%%%%%%%%%%%%%%%%%%   
\begin{table*}
\caption{The physical parameters of the 9 LMR EW systems from CSS. The second lines for CSS$\_$J210228.3-031048 and CSS$\_$J231513.3+345335 show the resulting physical parameters from the new multiband LCs.}
\label{tab:models}
   %\tiny
\scriptsize
   % \centering
\hspace{-1.4cm}
\begin{tabular}{ccccccccc}
\hline
 ID &  $q$ &  $\frac{T_{2}}{T_{1}}$ &  $\Omega_{1,2}$ &  $r_{1}$  & $r_{2}$  & $i (\degr)$ &   $\frac{L_{1}}{L_{\rm tot}}$   &  $f$  \\ \hline
 CSS$\_$J014859.6+391504 	&	0.16	$\pm$	0.02	&	0.940	$\pm$	0.012	&	2.085	$\pm$	0.009	&	0.558	$\pm$	0.037	&	0.254	$\pm$	0.044	&	82.9	$\pm$	2.8	&	0.865	&	0.43	$\pm$	0.09	\\
 CSS$\_$J065207.9+530125 	&	0.16	$\pm$	0.02	&	0.983	$\pm$	0.011	&	2.029	$\pm$	0.003	&	0.583	$\pm$	0.012	&	0.285	$\pm$	0.017	&	79.2	$\pm$	1.6	&	0.824	&	0.98	$\pm$	0.03	\\
 CSS$\_$J090748.9+375447 	&	0.10	$\pm$	0.02	&	1.046	$\pm$	0.013	&	1.897	$\pm$	0.004	&	0.610	$\pm$	0.021	&	0.244	$\pm$	0.030	&	79.6	$\pm$	3.1	&	0.843	&	0.96	$\pm$	0.07	\\
 CSS$\_$J101256.9+313709 	&	0.11	$\pm$	0.02	&	0.993	$\pm$	0.008	&	1.925	$\pm$	0.002	&	0.602	$\pm$	0.008	&	0.248	$\pm$	0.011	&	77.3	$\pm$	1.3	&	0.865	&	0.91	$\pm$	0.02	\\
 CSS$\_$J153254.1+342518 	&	0.09	$\pm$	0.02	&	0.986	$\pm$	0.011	&	1.922	$\pm$	0.011	&	0.591	$\pm$	0.048	&	0.202	$\pm$	0.060	&	83.6	$\pm$	2.6	&	0.900	&	0.09	$\pm$	0.18	\\
 CSS$\_$J174213.4+440857 	&	0.14	$\pm$	0.02	&	0.987	$\pm$	0.008	&	2.006	$\pm$	0.006	&	0.581	$\pm$	0.025	&	0.260	$\pm$	0.031	&	80.5	$\pm$	2.4	&	0.847	&	0.77	$\pm$	0.07	\\
 CSS$\_$J210228.3-031048 	&	0.13	$\pm$	0.02	&	1.017	$\pm$	0.007	&	2.021	$\pm$	0.011	&	0.569	$\pm$	0.045	&	0.234	$\pm$	0.054	&	79.9	$\pm$	5.4	&	0.848	&	0.31	$\pm$	0.13	\\
 	   &	0.15	$\pm$	0.01	&	1.011	$\pm$	0.001	&	2.030	$\pm$	0.002	&	0.574	$\pm$	0.010	&	0.259	$\pm$	0.011	&	79.4	$\pm$	4.0	&	0.830$^{a}$	&	0.57	$\pm$	0.03	\\
 CSS$\_$J224702.1+362815 	&	0.11	$\pm$	0.02	&	0.981	$\pm$	0.016	&	1.945	$\pm$	0.008	&	0.593	$\pm$	0.035	&	0.236	$\pm$	0.043	&	80.2	$\pm$	2.2	&	0.878	&	0.62	$\pm$	0.11	\\
 CSS$\_$J231513.3+345335 	&	0.12	$\pm$	0.02	&	1.044	$\pm$	0.010	&	1.953	$\pm$	0.012	&	0.594	$\pm$	0.055	&	0.251	$\pm$	0.071	&	79.7	$\pm$	4.3	&	0.826	&	0.84	$\pm$	0.16	\\
	  &	0.14	$\pm$	0.02	&	1.056	$\pm$	0.001	&	2.011	$\pm$	0.002	&	0.579	$\pm$	0.002	&	0.258	$\pm$	0.002	&	80.2	$\pm$	3.0	&	0.802$^{b}$	&	0.71	$\pm$	0.03	\\
\hline 
\end{tabular}
\tablenotetext{a} {In V$_J$ band. The $L_{1}/L_{\rm tot}$ in BRI bands are 0.827, 0.832, and 0.833, respectively.}
\tablenotetext{b}{In V$_J$ band. The $L_{1}/L_{\rm tot}$ in RI bands are 0.810, and 0.810, respectively.}
\end{table*}
%%%%%%%%%%%%%%%%%%%%%%%%%%%%%%%%%%%%%%%%%%%%%%%%%%%%%%%%%%%%%%%%%%%%%%
\begin{table*}
\caption{The absolute parameters of the 9 LMR EW systems from CSS.}\label{tab:absparams}
\scriptsize
    %\centering
\hspace{-2.2cm}
\begin{tabular}{cccccccccc}
 \hline
 ID  &  $T_{\rm sys}$ &   $T_{1}$  &   $T_{2}$ &  $M_{1}$   & $M_{2}$ & $R_{1}$ &  $R_{2}$  & $L_{1}$  &  $L_{2}$  \\
&  (K) & (K) & (K) &  ($M_{\sun}$) &  ($M_{\sun}$)  & ($R_{\sun}$) & ($R_{\sun}$) & ($L_{\sun}$) & ($L_{\sun}$) \\ 
 \hline 
  CSS$\_$J014859.6+391504 	&	7045	$\pm$	271	&	7113	$\pm$	274	&	6686	$\pm$	257	&	2.06	$\pm$	0.15	&	0.33	$\pm$	0.05	&	2.66	$\pm$	0.19	&	1.21	$\pm$	0.21	&	20.38	$\pm$	6.62	&	2.62	$\pm$	1.00	\\
  CSS$\_$J065207.9+530125 	&	7262	$\pm$	63	&	7286	$\pm$	63	&	7162	$\pm$	62	&	1.73	$\pm$	0.05	&	0.28	$\pm$	0.04	&	2.58	$\pm$	0.06	&	1.26	$\pm$	0.08	&	8.72	$\pm$	0.61	&	3.74	$\pm$	0.47	\\
  CSS$\_$J090748.9+375447 	&	5625	$\pm$	112	&	5588	$\pm$	111	&	5845	$\pm$	116	&	1.11	$\pm$	0.03	&	0.11	$\pm$	0.02	&	1.25	$\pm$	0.05	&	0.50	$\pm$	0.06	&	1.04	$\pm$	0.13	&	0.26	$\pm$	0.07	\\
  CSS$\_$J101256.9+313709 	&	6295	$\pm$	76	&	6301	$\pm$	76	&	6257	$\pm$	75	&	1.38	$\pm$	0.03	&	0.15	$\pm$	0.03	&	1.49	$\pm$	0.03	&	0.61	$\pm$	0.03	&	3.03	$\pm$	0.20	&	0.52	$\pm$	0.05	\\
  CSS$\_$J153254.1+342518 	&	6747	$\pm$	45	&	6757	$\pm$	45	&	6662	$\pm$	44	&	1.67	$\pm$	0.05	&	0.15	$\pm$	0.03	&	1.80	$\pm$	0.15	&	0.61	$\pm$	0.18	&	7.46	$\pm$	0.37	&	0.67	$\pm$	0.40	\\
  CSS$\_$J174213.4+440857 	&	6331	$\pm$	178	&	6345	$\pm$	178	&	6262	$\pm$	176	&	1.50	$\pm$	0.04	&	0.21	$\pm$	0.03	&	1.65	$\pm$	0.07	&	0.74	$\pm$	0.09	&	4.45	$\pm$	0.31	&	0.75	$\pm$	0.20	\\
  CSS$\_$J210228.3-031048 	&	6140	$\pm$	209	&	6129	$\pm$	209	&	6195	$\pm$	211	&	1.16	$\pm$	0.03	&	0.17	$\pm$	0.01	&	1.04	$\pm$	0.02	&	0.47	$\pm$	0.02	&	1.28	$\pm$	0.16	&	0.29	$\pm$	0.05	\\
  CSS$\_$J224702.1+362815 	&	6333	$\pm$	142	&	6349	$\pm$	142	&	6228	$\pm$	140	&	1.55	$\pm$	0.05	&	0.17	$\pm$	0.03	&	1.85	$\pm$	0.11	&	0.74	$\pm$	0.14	&	5.24	$\pm$	0.67	&	0.73	$\pm$	0.28	\\
  CSS$\_$J231513.3+345335 	&	5548	$\pm$	145	&	5493	$\pm$	144	&	5803	$\pm$	152	&	1.02	$\pm$	0.04	&	0.14	$\pm$	0.01	&	1.06	$\pm$	0.01	&	0.47	$\pm$	0.01	&	0.71	$\pm$	0.13	&	0.23	$\pm$	0.02	\\

\hline
\end{tabular}
\end{table*}
%%%%%%%%%%%%%%%%%%%%%%%%%%%%%%%%%%%%%%%%%%%%%%%%%%%%%%%%%%%%%%%
%%%%%%%%%%%%%%%%%%%%%%%%%%%%%%%%%%%%%%%%%%%%%%%%%%%%%%%%%%%%%%%%%%

%%%%%%%%%%%%%%%%%%%%%%%%%%%%%%%%%%%%%%%%%%%%%%%%%%%%%%%%%%
%%%%%%%%%%%%%%%%%%%%%%%%%%%%%%%%%%%%%%%%%%%%%%%%%%%%%%%%%%
%%%%%%%%%%%%%%%%%%%%%%%%%%%%%%%%%%%%%%%%%%%%%%%%%%%%%%%%%
%%%%%%%%%%%%%%%%%%%%%%%%%%%%%%%%%%%%%%%%%%%%%%%%%%%%%%%%%
%%%%%%%%%%%%%%% physicalparams%%%%%%%%%%%%%%%%%%%%%%%%%%%%%%%%%%%%%
\begin{figure*}
  \minipage{0.5\textwidth}
  \centering
  \includegraphics[width=\linewidth]{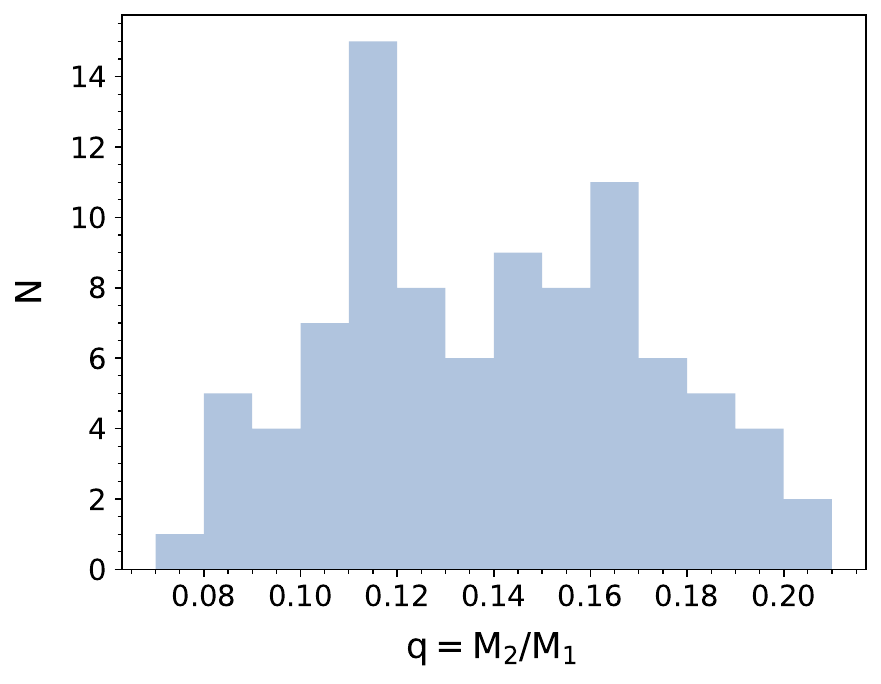} %
  \center \textbf{(a)}
  \endminipage%\hfill
  \minipage{0.5\textwidth}
  \includegraphics[width=\linewidth]{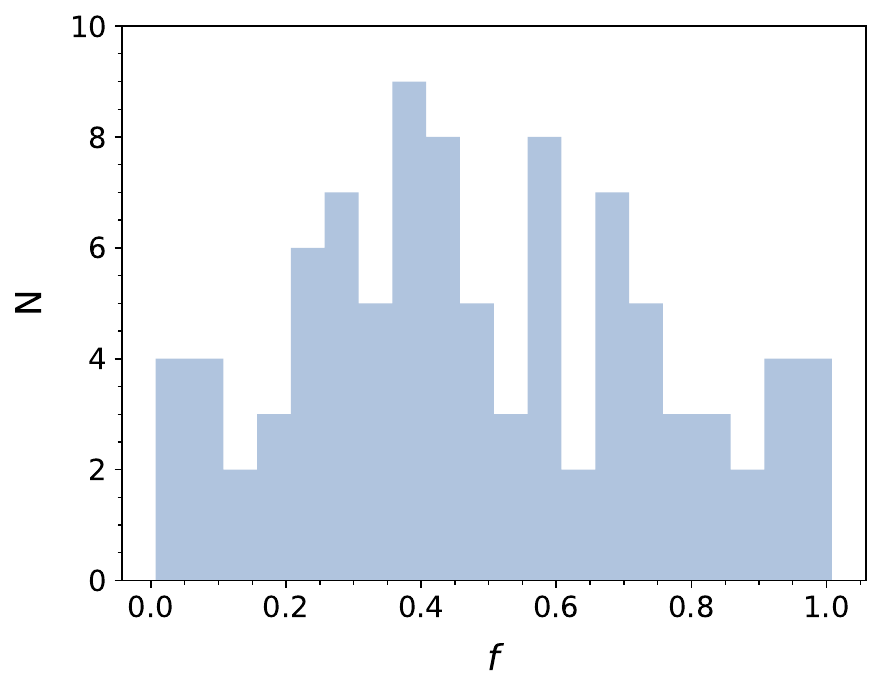} %
  \center \textbf{(b)}
  \endminipage
  \vspace{+0.3cm}

 \minipage{0.5\textwidth}
 \centering
 \includegraphics[width=\linewidth]{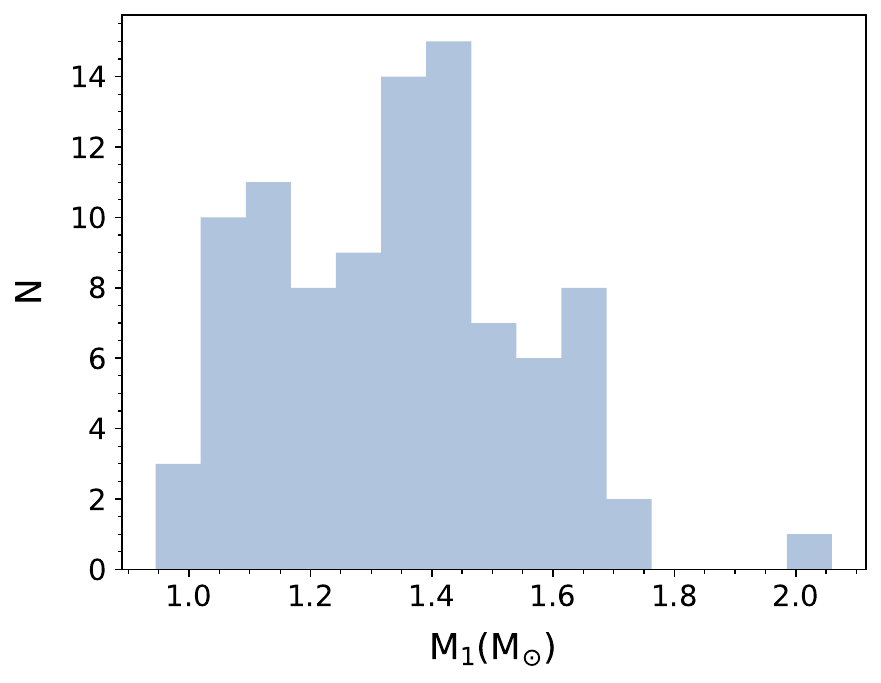} %
 \center \textbf{(c)}
 \endminipage%\hfill
 \minipage{0.5\textwidth}
 \includegraphics[width=\linewidth]{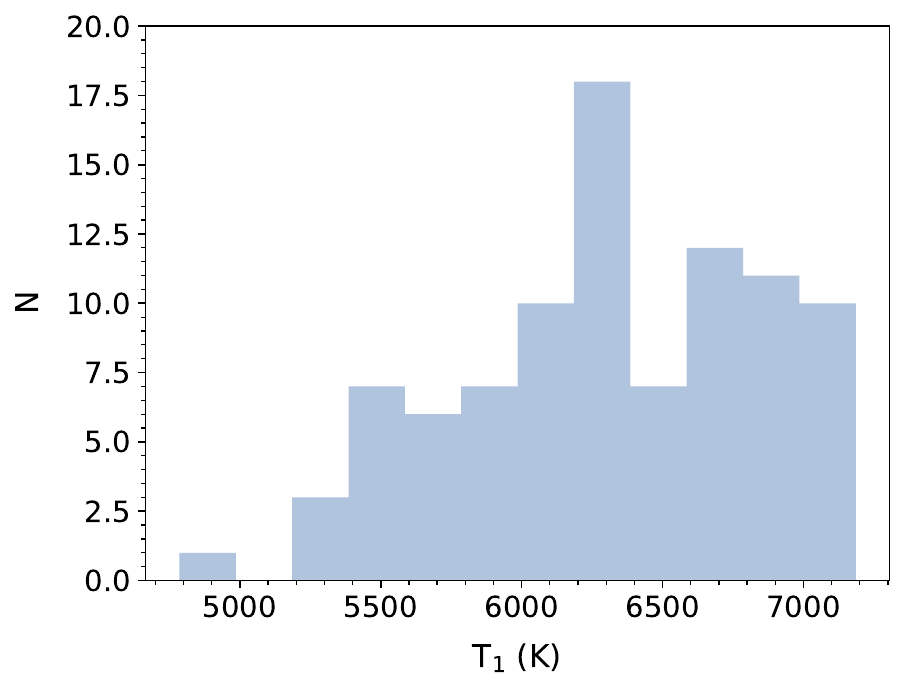} %
 \center \textbf{(d)}
 \endminipage
 \vspace{+0.3cm}
 \caption{The distributions of \textbf{(a)} $q$, \textbf{(b)} $f$, \textbf{(c)} $M_{1}$ and \textbf{(d)} $T_{1}$ of the 92 CSS LMRs.}\label{fig:qfM1T192}
\end{figure*}
%%%%%%%%%%%%%%%%%%%%%AbsParams%%%%%%%%%%%%%%%%%%%%%%%%% %%%%%%%%%%%%%%%%%%%%%%%%%%%%%%%%%%%%%%%%%%%%%%%%%%%%%%%%%%%%
\begin{figure*}
 \minipage{0.45\textwidth}
 \centering
 \includegraphics[width=\linewidth]{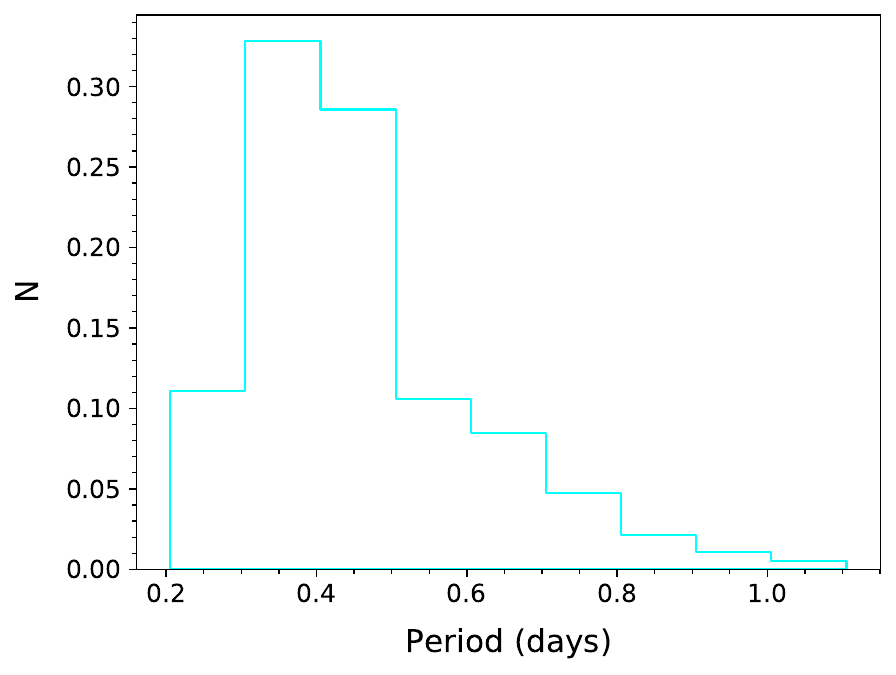} %
 \center \textbf{(a)}
 \endminipage
 \hspace{+1.5cm}
 \minipage{0.45\textwidth}
 \includegraphics[width=\linewidth]{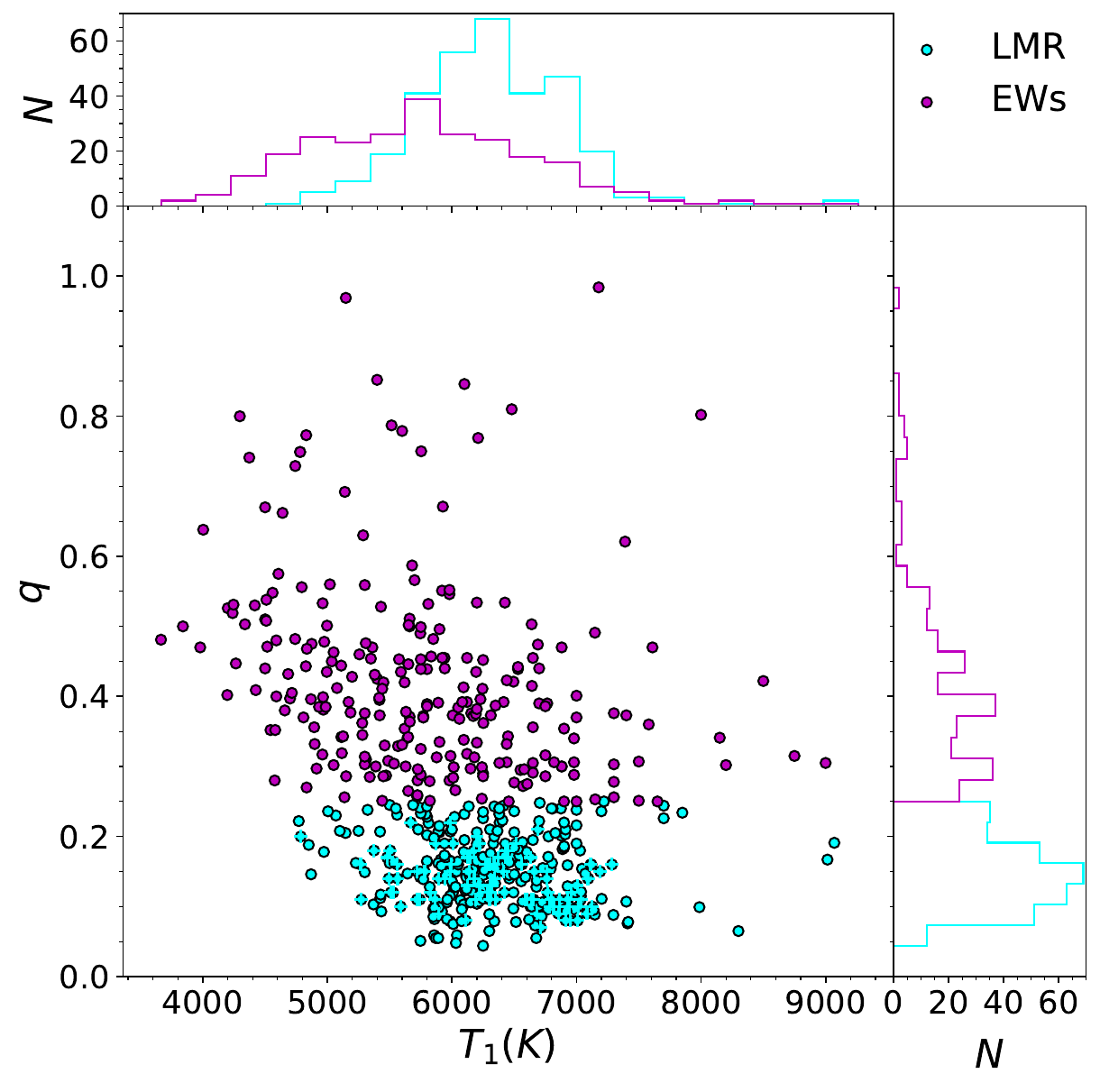} %
 \center \textbf{(b)}
 \endminipage
 \vspace{+0.3cm}

 \minipage{0.45\textwidth}
 \centering
 \includegraphics[width=\linewidth]{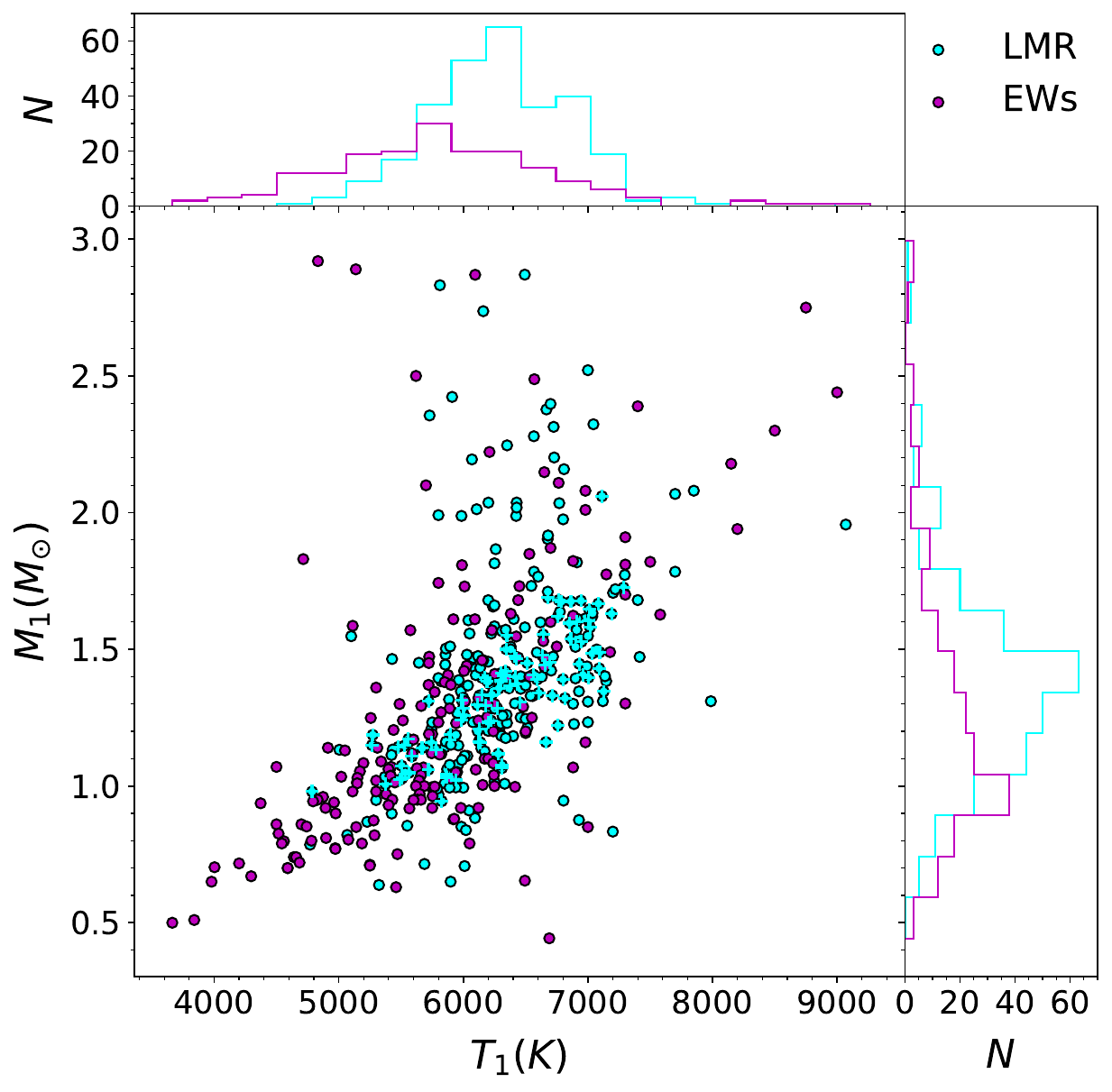} %
 \center \textbf{(c)}
 \endminipage%\hfill
 \hspace{+1.5cm}
 \minipage{0.45\textwidth}
 \includegraphics[width=\linewidth]{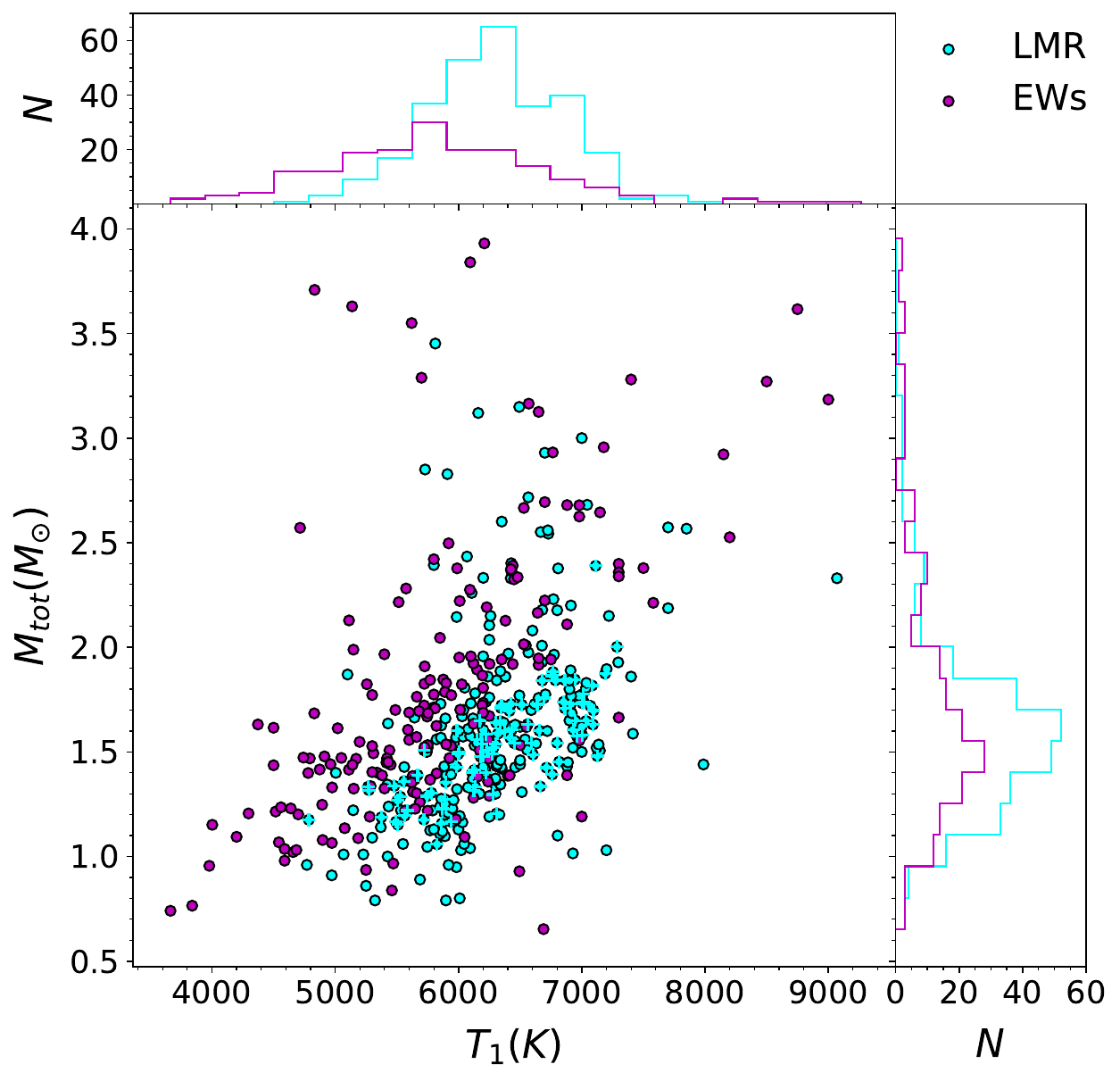} %
 \center \textbf{(d)}
 \endminipage
 \vspace{+0.3cm}
 \caption{(a) The period distribution of the 92 LMRs from CSS and the 98 LMRs from ASAS-3. (b-d) Dependence of $T_{1}$ on $q$, $M_{1}$, and $M_{\rm tot}$  respectively, for the sample of 317 LMRs (cyan) and the 253 well-studied EWs \citep{2021PASJ...73..132L} (purple). The upper panel in each plot represents the corresponding distributions of $T_{1}$ for the two samples whereas the right-hand histograms in (c) and (d) represent the distribution of $M_{1}$ and $M_{\rm tot}$  respectively for the two samples.}\label{fig:ALL}
\end{figure*}
%%%%%%%%%%%%%%%%%%%%%%%%%%%%%%%%%%%%%%%%%%%%%%%%%%
\begin{figure*}
 \vspace{+0.3cm}
 \minipage{0.5\textwidth}
 \centering
 \includegraphics[width=\linewidth]{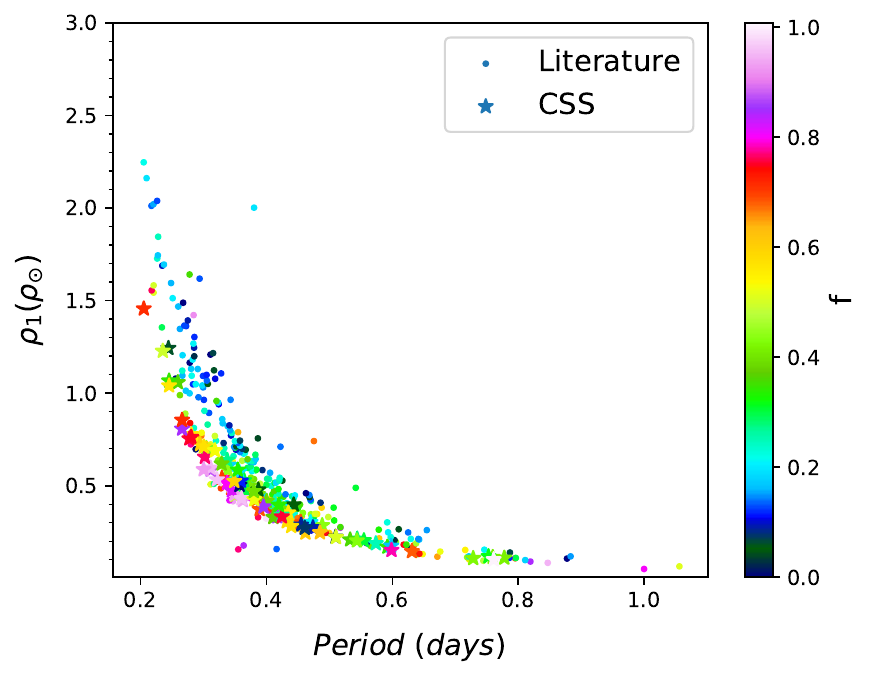} %
 \center \textbf{(a)}
 \endminipage
 \minipage{0.5\textwidth}
 \includegraphics[width=\linewidth]{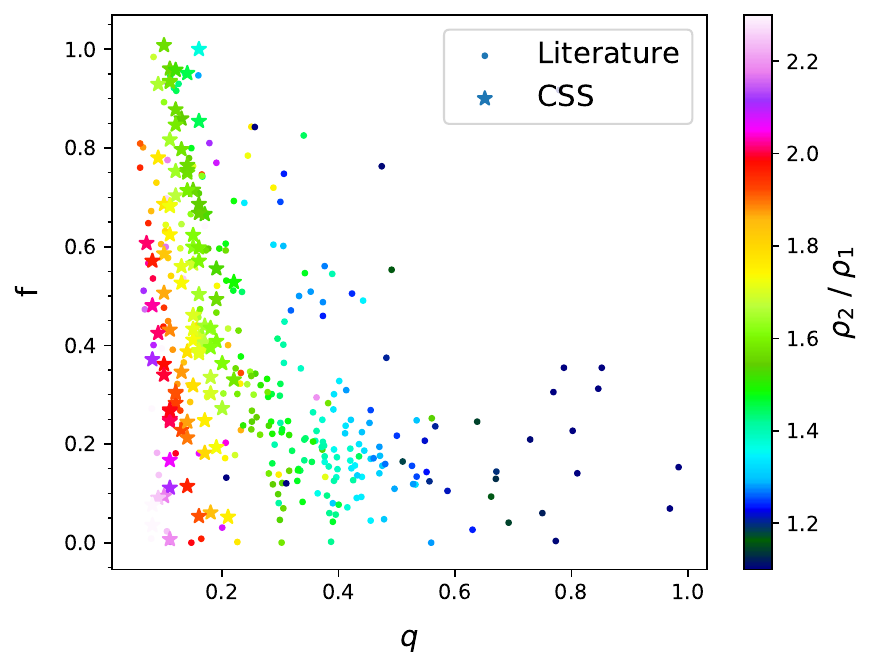} %
 \center \textbf{(b)}
 \endminipage
 \vspace{+0.3cm}
 \minipage{0.5\textwidth}
 \centering
 \includegraphics[width=\linewidth]{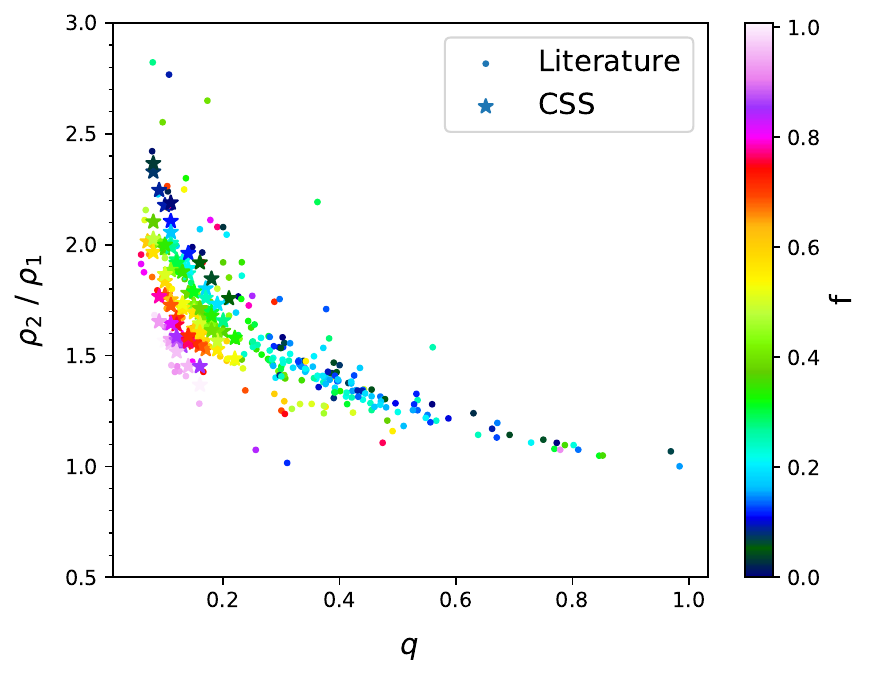}
 \center \textbf{(c)}
 \endminipage
 \vspace{+0.3cm}
 \caption{(a) The mean density of the primary component $\rho_{1} (\rho_{\sun})$ as a function of period and fillout factor $f$. (b) The mass ratio $q$ as a function of fillout factor $f$ and the ratio of densities $\rho_2/\rho_1$. (c) The ratio of densities $\rho_2/\rho_1$ of the components as a function of mass ratio $q$ and fillout factor $f$. Color-coded according to the scale on the right.}\label{fig:densities}
\end{figure*}
%%%%%%%%%%%%%%%%%%%%%%%%%%%%%%%%%%%%%%%
%%%%%%%%%%%%%%%%%%%%%%%%%%%%%%%%%%%%%%%%%%%%%%%%%%%%%%%%%%%%%
\begin{figure*}
 \minipage{0.5\textwidth}
 \centering
 \includegraphics[width=\linewidth]{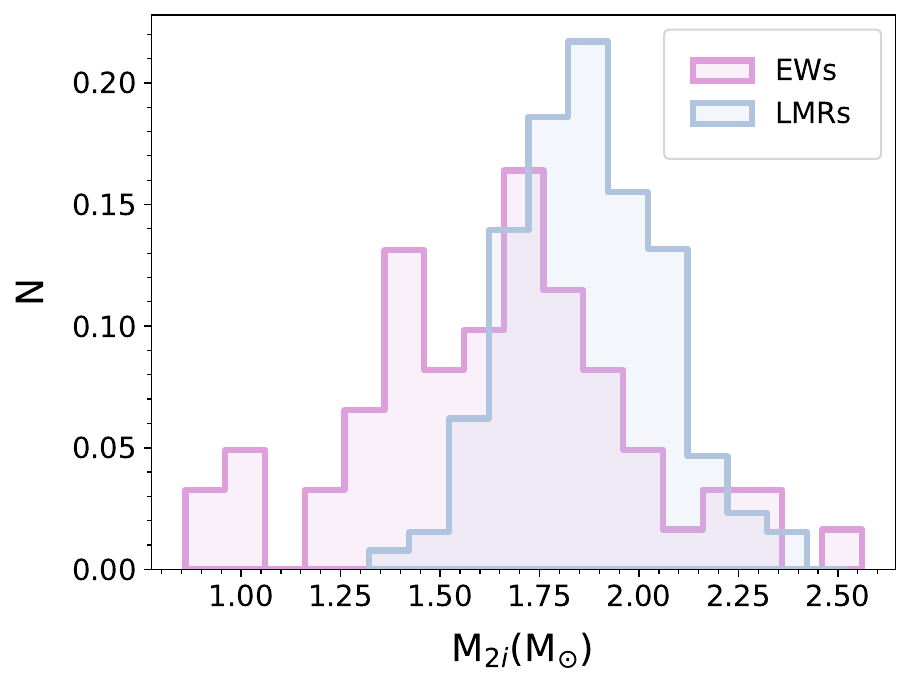} %
 \center \textbf{(a)}
 \endminipage%\hfill
 \minipage{0.5\textwidth}
 \includegraphics[width=\linewidth]{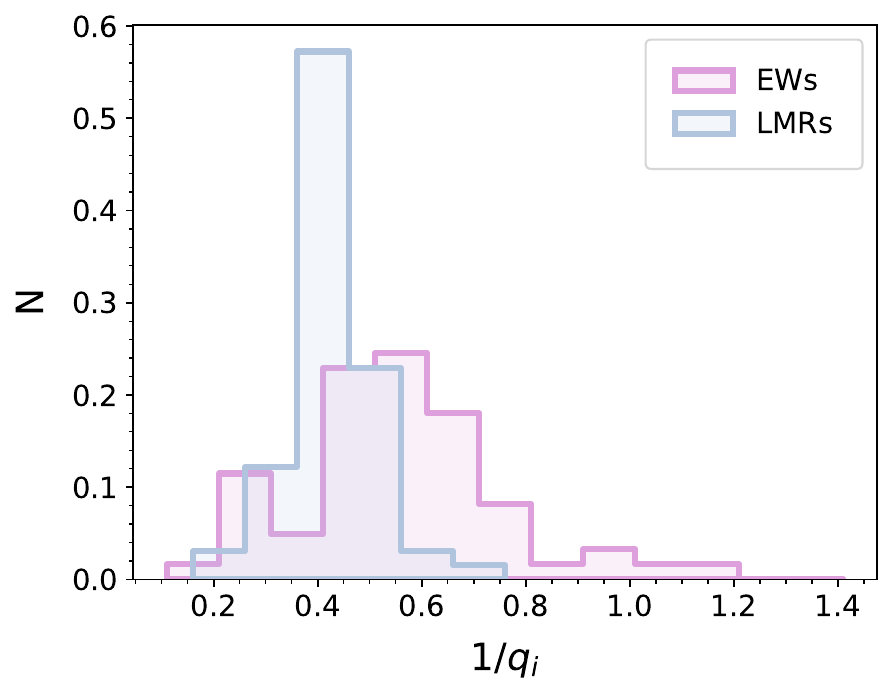} %
 \center \textbf{(b)}
 \endminipage
 \vspace{+0.3cm}
 \caption{(a) The mass distribution of the progenitors of the initial massive component ($M_{2\rm i}$) and (b) the distribution of the initial mass ratio ($1/q_{\rm i}$) of the progenitors of LMRs from CSS and \citet{2013MNRAS.430.2029Y} and EWs with $q>0.25$ from the latter.}\label{fig:progeni}%from CSS and \citet{2013MNRAS.430.2029Y} and EWs with $q>0.25$ from the latter.}
\end{figure*}
%\section{Results} \label{sec:Results}

\section{Discussion and Conclusions} \label{sec:Discussion}
We present the identification and photometric investigation of 7 new LMR totally eclipsing contact binaries discovered in the CSS as well as multiband photometry and analysis of two totally eclipsing LMRs included in the automatic photometric analysis by \cite{2020ApJS..247...50S} of 2335 late-type contact binaries from CSS. All 7 were found to be of extremely low mass ratio, ranging from 0.09 to 0.16.

In our first application of the Fourier LMR discovery project to the CSS survey, we identified 92 systems \citep{2020CoSka..50..409L}. Among these, 37 systems were previously unknown \cite[][and current work]{2022MNRAS.512.1244C}, while the remaining 55 were previously reported by \cite{2020ApJS..247...50S}. For our analysis, we focused on the sample of 92 LMRs, ensuring consistency in methodology and assumptions to perform limited statistics. As shown in Fig.~\ref{fig:qfM1T192}, the mass ratio distribution of the 92 CSS LMR systems has a peak at 0.12, although a significant number of systems has $q=0.17$. %10$\%$ of the systems have $q<0.10$ while the lower value is 0.07$\pm$ 0.02 (J075839).%  
In 10$\%$ of the systems, $q$ is less than 0.10, with the lowest recorded value being 0.07$\pm$ 0.02 (J075839).
The primary masses $M_{1}$ are in the range of $1.0-1.4\, M_{\sun}$, whereas there are many systems with $M_{1}>1.4\, M_{\sun}$, which is unusual for EW systems (e.g., 22$\%$ of EW systems have $M_{1}>1.5\, M_{\sun}$). The distribution of $T_{1}$ shows that the majority has $T_{1}>6000~ K,$ and 13\% have $T_{1}>7000~ K,$ which is unusual given the common range of 5500-6000~K of EW systems \citep{2021ApJS..254...10L}. %The 92 LMRs present deep, medium, and shallow degrees of contact, although 50\% of the systems have $f>45\%$.
\textbf{Half of the 92 LMRs have $f>45\%$ and within the uncertainties of the fill out parameter, belong to deep contact systems.}
Fig.~\ref{fig:ALL}a shows a histogram of the period distribution for our sample of 92 CSS LMRs and the 98 new LMR candidates from ASAS-3 (see Appendix~\ref{sub:ASAS-3}). There are 20\% of systems with $P>0.5$~d.

To compare the 92 LMRs with the rest of the EWs, we also included 173 well-studied ($q$ from spectroscopy or/and total eclipses) LMRs from the literature from the catalog of \cite{2022MNRAS.512.1244C}, updated up to 2023 with $\sim$50 more systems (total 317 LMRs) with $0.05<q<0.25$, and the 253 well-studied EWs from \cite{2021ApJS..254...10L} with $q>0.25$ (so as not to overlap with the LMRs). The dependence of $T_{1}$ on $q$, $M_{1}$, and $M_{\rm tot}$ for the two samples is presented in Fig.~\ref{fig:ALL}b, c, and d, respectively. The histogram on the upper panel in each plot represents the distribution of $T_{1}$ for each sample, whereas the right-hand histograms in (c) and (d) represent the distribution of $M_{1}$ and $M_{\rm tot}$ respectively, for the two samples. Note that the histogram of $T_{1}$ in Fig.~\ref{fig:ALL}b is slightly different from that of Fig.~\ref{fig:ALL}c and d, as no absolute parameters have been determined for all LMRs and EWs.
As displayed in Fig.~\ref{fig:ALL}c, LMR systems show a trend to have warmer (top histogram) and more massive (right histogram) primaries.
 
To further investigate the properties of the LMRs, we calculated the mean densities of the primary and secondary components, $\rho_{1,2}$, for both the 92 CSS and the literature systems \citep[from][]{2021ApJS..254...10L}. The results show that systems with shorter periods tend to consist of components with larger $\rho_{1}$, while for periods shorter than 0.4~days, systems with the same period have a higher $f$ for smaller $\rho_1$ (Fig.~\ref{fig:densities}a). We also find that systems with higher $f$ reach lower $\rho_2/\rho_1$ towards merging (for $q$ less than 0.10) and that the ratio $\rho_2/\rho_1$ of systems with the same mass ratio follows a downward trend as the fillout factor $f$ increases (Figs.~\ref{fig:densities}b and c, respectively). 

To investigate the dynamical stability of all 9 systems with mass ratios that are extremely close to the minimum mass ratio for contact systems, we calculate the ratio of the spin angular momentum ($J_{\rm s}$) to the orbital angular momentum ($J_{\rm o}$) using the formula derived from equations (1) and (2) of \cite{2006MNRAS.369.2001L}, namely
\begin{equation}
 \frac{J_{\rm s}}{J_{\rm o}}= \frac{(1+q)}{q} (k_{1} r_{1})^{2} \left[1+q \left(\frac{k_{2}}{k_{1}}\right)^{2} \left(\frac{r_{2}}{r_{1}}\right)^2  \right],
\label{eq:2}
\end{equation}
\noindent where $k_{1}, k_{2}$ are the dimensionless gyration radii of both components. As the value of $k$, which determines the interior structure of the star, is the primary source of uncertainty, we initially adopt $k_{1}^2= k_{2}^2=k^2=0.06$ (Sun-like) as in \cite{1995ApJ...444L..41R}. The ratio values $J_{\rm s}/J_{\rm o}$ for the 9 LMR systems %are shown in the second column of Table~\ref{tab:table4} and 
are all below 1/3, indicating that all systems are stable. 
The ratio was then adjusted using the value of $k_{2}^2=0.205$ \citep{2007MNRAS.377.1635A} for a convective secondary due to its very low mass. The value of $k_{1}$ was derived from the linear $k_{1} - M$ relationships provided by \citet{2022MNRAS.512.1244C}, which are based on the values tabulated by \citet{2009A&A...494..209L}. Specifically, we use $k_{1}=-0.250 \, M+0.539$ for stars with $M=0.5-1.4 \, M_{\sun}$, and $k_{1}=0.014 \, M+0.152$ for stars with $M>1.4\,M_{\sun}$.
%The results are listed in the third column of Table~\ref{tab:table4}.
There are only two systems with a  $(J_{\rm s}/J_{\rm o})_{\rm k}$ ratio close to 1/3, J090748 (0.30) and J231513 (0.28). Nevertheless, for J231513 the ratio $(J_{\rm s}/J_{\rm o})_{\rm k}$ reduces to 0.24 when we use the results of the spotted solution from our VRI LCs. The non-uniqueness of the spotted solutions is an old issue in the modeling of contact binaries %It should be noted that modeling the system by introducing spots makes the solution non-unique and 
so there are potentially other solutions that may show that systems such as J231513 are stable. Another issue with spotted solutions %In addition, we have to point out 
is that the LCs of contact binaries at different passbands of the surveys (ASAS-SN, ZTF, TESS) and of the dedicated observations \citep{2023AJ....165...80P}, do not always show the same asymmetry (variable O'Connell effect) and this may alter the geometric solution and subsequently the absolute parameters. 
  
The degenerate solutions  (different but equally well fitting spotted light curve solutions) in the LC modeling, lead to the degeneracy of instability parameters such as the instability mass ratio $q_{\rm inst}$ which depends on the mass of the primary component and the fillout factor \citep{2021MNRAS.501..229W} and/or the metallicity \citep{2024MNRAS.527....1W}. This highlights the need to find a different instability criterion in the absence of Doppler Imaging analysis and high dispersion spectra. 

As interesting as the end of an LMR's life is, so are their ancestors. Even though many questions remain regarding the specific evolutionary process through the detached channel, the initial masses of the two components play a critical role in the evolution to the contact phase. 
\cite{2013MNRAS.430.2029Y} proposed a scenario that allows us to calculate these values, based on the findings from the mass-luminosity and mass-radius diagrams. In this framework, primary components of EWs reside in the region of unevolved low-mass single stars between the zero-age main sequence (ZAMS) and terminal-age main sequence (TAMS), while secondaries are systematically oversized and overluminous for their masses. Assuming a reversal of roles and mass transfer, with both components expanding beyond their Roche lobes and sharing a common envelope, following the above method, we calculated the initial mass of the primary ($M_{1\rm i}$) and the secondary component ($M_{2\rm i}$) as well as the initial mass ratio ($1/q_{\rm i}$) for the 92 CSS LMRs.
The results of the above procedure on the 92 CSS LMRs show that the mean mass of the progenitors is $M_{2\rm i}=1.87\pm0.21 \, M_\odot$ and $M_{1\rm i}= 0.78\pm0.16 \, M_\odot$. As is shown in Fig.~\ref{fig:progeni}a, the distribution of the masses of the progenitors of the initial massive components ($M_{2\rm i}$) of our LMRs and EW sample \citep[62 spectroscopic studied EWs with $q>0.25$ from][]{2013MNRAS.430.2029Y} reveals that LMR systems tend to originate from higher-mass ancestors. This could imply that, similarly to single stars, their evolution is mainly dictated by the thermonuclear reactions taking place in their cores. In the case of ultra-short-period EWs, on the other hand, evolution is instead primarily regulated by AML, mass loss, and magnetic braking \citep{2023AJ....165...80P}.

In order to achieve our initial objective of applying the same technique for identification, photometric analysis, modeling, and extraction of absolute parameters to a homogeneous sample (CSS) of LMRs for further quantitative conclusions and discussion of mini statistics, it is important to clarify that we did not use data from TESS (where available). Future research on the most intriguing LMRs will involve a period variation study as well as TESS analysis together with multiband ZTF (gri), ASAS-SN light curves, and other surveys as in \cite{2023AJ....165...80P}.

\begin{acknowledgments}
EL gratefully acknowledges the support provided by IKY ``Scholarship Programme for PhD candidates in the Greek Universities''. AP gratefully acknowledges the support provided by the grant co-financed by Greece and the European Union (European Social Fund- ESF) through the Operational Programme «Human Resources Development, Education and Lifelong Learning» in the context of the project “Reinforcement of Postdoctoral Researchers-2nd Cycle” (MIS-5033021), implemented by the State Scholarships Foundation (IKY). Support for C.E.F.L. and M.C. is provided by ANID’s FONDECYT Regular grant 1231637 and Millennium Science Initiative grant ICN12\_009, awarded by ANID to the Millennium Institute of Astrophysics (MAS). Additional support for M.C. is provided by ANID's Basal grant FB210003. \\
\end{acknowledgments}

\hspace{5mm}
\facilities{Based on observations made with the 2.3 m Aristarchos telescope, Helmos Observatory, Greece, which is operated by the Institute for Astronomy, Astrophysics, Space applications and remote sensing of the National Observatory of Athens, Greece.
Catalina Sky Survey.}


\software{\textsc{PHOEBE-0.31a} \citep{2005ApJ...628..426P}, 
          \textsc{MWDUST} \citep{2016ApJ...818..130B},
          \textsc{PyRAF} \citep{2012ascl.soft07011S},
          \textsc{Astrometry.net} \citep{2010AJ....139.1782L},      
          }

\appendix
\section{New LMR candidates from ASAS-3}\label{sub:ASAS-3}
We apply the same method as described by \cite{2020CoSka..50..409L} and \cite{2022MNRAS.512.1244C} on the eclipsing binary catalog of All Sky Automated Survey \citep[ASAS-3,][]{ASAS-3,2002AcA....52..397P}
and identified 98 LMR candidates. Table~\ref{tab:Table5} lists the reference time of minimum $\rm HJD_0$, the Period (days), and the ASAS-3 magnitude $V_{\rm ASAS}$ (mag). Approximately 21 out of the 98 candidates are confirmed LMRs in previous studies (denoted by $\alpha$ in Table~\ref{tab:Table5}) and 4 are potential red nova progenitors according to \cite{2022JApA...43...94W} (denoted by $b$ in Table~\ref{tab:Table5}). Together, the 37 new LMRs from CSS and the 77 LMR candidates from ASAS-3 provide an excellent chance for ground-based dedicated observations to further whittle down the list of possible mergers.
\begin{table*}
% \centering
 \caption{The 98 LMR candidates from ASAS-3.}
 \label{tab:Table5}
 \hspace{-2.4cm}
 \begin{tabular}{lccclccc}
 \hline
 \multicolumn{1}{c}{\text{Name}} & $\rm HJD_0$ & Period & $V_{\rm ASAS}$ & \multicolumn{1}{c}{\text{Name}} & $\rm HJD_0$ & Period & $V_{\rm ASAS}$  \\ 
 & ($2450000+$) & (days) & (mag) & & ($2450000+$) & (days) & (mag) \\
 \hline
ASAS$\_$J	001556+0644.7	&	3169.92649	&	0.4012003	&	11.17		&	ASAS$\_$J	101159+1652.5$^{a}$	&	2667.77002	&	0.2667329	&	11.46	\\
ASAS$\_$J	004717-1941.6	&	3185.86545	&	0.4888105	&	11.31		&	ASAS$\_$J	101818-5154.6	&	2634.81936	&	0.5846602	&	10.29	\\
ASAS$\_$J	014255-2007.5	&	2460.98228	&	0.3659550	&	11.11		&	ASAS$\_$J	102556+2049.3	&	4476.83060	&	0.2849765	&	10.43	\\
ASAS$\_$J	015937-0331.0	&	2644.55908	&	0.6315200	&	9.35		&	ASAS$\_$J	103231+1608.6	&	2689.73888	&	0.6456704	&	10.63	\\
ASAS$\_$J	023152-3837.4	&	4400.65914	&	0.5887406	&	9.91		&	ASAS$\_$J	105210-4345.5	&	3403.80885	&	0.6104604	&	10.91	\\
ASAS$\_$J	032629-3101.3	&	1876.61098	&	0.6386700	&	10.81		&	ASAS$\_$J	111131-0815.7	&	2740.63563	&	0.6008158	&	9.79	\\
ASAS$\_$J	033142-5927.4	&	2921.83140	&	0.3779974	&	12.17		&	ASAS$\_$J	122119-1359.9$^{a}$	&	1962.73746	&	0.4745019	&	10.98	\\
ASAS$\_$J	034202-5145.2	&	3631.78155	&	0.4092128	&	10.82		&	ASAS$\_$J	124338-4656.1	&	2831.59247	&	0.6693620	&	10.23	\\
ASAS$\_$J	034809-5839.8	&	3798.52851	&	0.4292427	&	10.48		&	ASAS$\_$J	125736+0749.2	&	2804.56497	&	0.3636122	&	11.61	\\
ASAS$\_$J	035020-8017.4$^{a}$	&	4553.49053	&	0.6224005	&	11.94		&	ASAS$\_$J	130226+0718.6	&	3186.58347	&	0.4080725	&	11.70	\\
ASAS$\_$J	035200-2155.8	&	2136.81948	&	0.3351662	&	10.69		&	ASAS$\_$J	135314+2009.7	&	3129.66797	&	0.5315500	&	10.34	\\
ASAS$\_$J	040315+1411.5	&	2987.70073	&	0.3803151	&	11.79		&	ASAS$\_$J	143652-6646.6	&	3113.76872	&	0.7020080	&	10.44	\\
ASAS$\_$J	040528-6536.2	&	3767.69194	&	0.2947729	&	11.91		&	ASAS$\_$J	144235-4027.2$^{a}$	&	1996.73775	&	0.3250320	&	10.83	\\
ASAS$\_$J	040550-5402.6	&	3792.74450	&	0.3601228	&	11.51		&	ASAS$\_$J	145538-7948.4$^{a}$	&	1980.73633	&	1.0655580	&	9.13	\\
ASAS$\_$J	041138-4438.0	&	3437.53380	&	0.8945652	&	9.42		&	ASAS$\_$J	153708-0606.3	&	2879.53968	&	0.9179296	&	10.67	\\
ASAS$\_$J	045211-2511.7	&	3407.62308	&	0.5789031	&	11.41		&	ASAS$\_$J	155025-0757.3	&	3901.70901	&	0.3578460	&	11.72	\\
ASAS$\_$J	045707-7207.9	&	2962.71682	&	0.4183947	&	12.08		&	ASAS$\_$J	160847+2511.7	&	4229.60604	&	0.3502237	&	11.60	\\
ASAS$\_$J	050334-2521.9$^{a}$	&	2197.76363	&	0.4140682	&	11.09		&	ASAS$\_$J	170715-5118.7$^{a}$	&	2834.76161	&	0.5258890	&	11.40	\\
ASAS$\_$J	051306+1558.2$^{a}$	&	3266.86384	&	0.3830040	&	11.75		&	ASAS$\_$J	171905-6313.2	&	2068.65499	&	0.4547656	&	10.71	\\
ASAS$\_$J	052650-8135.2	&	4216.46844	&	0.4616655	&	8.17		&	ASAS$\_$J	173400+1614.0	&	3273.58466	&	0.9736282	&	11.07	\\
ASAS$\_$J	055624-5919.5	&	3653.81800	&	0.6191750	&	11.88		&	ASAS$\_$J	173638-6648.1	&	2502.57297	&	0.4419562	&	11.19	\\
ASAS$\_$J	055827-1739.8	&	3430.64794	&	0.4145902	&	10.15		&	ASAS$\_$J	175656-3055.3	&	4540.88949	&	0.3952890	&	11.67	\\
ASAS$\_$J	060011-1549.8	&	4336.87729	&	0.4169109	&	12.02		&	ASAS$\_$J	180157-7228.1	&	4204.86185	&	0.3559101	&	10.48	\\
ASAS$\_$J	061627-7426.8	&	3779.68621	&	0.6304502	&	10.63		&	ASAS$\_$J	180433-4213.4	&	3832.85690	&	0.7008844	&	8.22	\\
ASAS$\_$J	061717-3427.6	&	3017.58607	&	0.5344865	&	11.02		&	ASAS$\_$J	181003-8126.2	&	4315.71549	&	0.3815023	&	11.56	\\
ASAS$\_$J	061758-0714.2	&	3433.69821	&	1.1280524	&	9.40		&	ASAS$\_$J	183317-4208.7	&	2879.16287	&	0.4566467	&	11.59	\\
ASAS$\_$J	061911-1548.3	&	3826.49788	&	0.4109543	&	10.47		&	ASAS$\_$J	184644-2736.4$^{a,b}$	&	2863.48857	&	0.3028361	&	11.65	\\
ASAS$\_$J	063420-4833.3	&	2258.70701	&	0.3215688	&	9.16		&	ASAS$\_$J	185354-3454.0	&	2795.77601	&	0.3946822	&	10.82	\\
ASAS$\_$J	063546+1928.6$^{a}$	&	4433.71624	&	0.4755129	&	9.95		&	ASAS$\_$J	190704+0742.0	&	2561.51030	&	0.8136180	&	10.76	\\
ASAS$\_$J	065227-5524.6	&	2244.77458	&	0.7228700	&	11.14		&	ASAS$\_$J	191205-4013.7	&	2909.66317	&	0.6457323	&	11.79	\\
ASAS$\_$J	065357-3648.1	&	3801.55019	&	0.3983703	&	10.84		&	ASAS$\_$J	192111-2228.9	&	4163.87647	&	0.8123023	&	11.49	\\
ASAS$\_$J	065638-3850.1	&	2213.76269	&	0.4257200	&	10.36		&	ASAS$\_$J	192349+0818.4$^{a}$	&	3467.88454	&	0.4237983	&	9.12	\\
ASAS$\_$J	070140-1514.1	&	3363.68534	&	0.5885460	&	11.78		&	ASAS$\_$J	192626-3019.2	&	4400.57511	&	0.4469845	&	10.37	\\
ASAS$\_$J	071451-5916.1$^{a}$	&	3511.45194	&	0.4722930	&	9.17		&	ASAS$\_$J	193822-0332.6$^{a}$	&	3159.75451	&	0.4127745	&	10.86	\\
ASAS$\_$J	071603-3052.1	&	2755.53485	&	0.3699650	&	11.26		&	ASAS$\_$J	194512-1932.8	&	3851.89999	&	0.6728200	&	10.92	\\
ASAS$\_$J	072525+1052.1	&	3795.65301	&	0.5401389	&	11.28		&	ASAS$\_$J	194625+0845.1$^{a}$	&	2861.61954	&	0.4175379	&	11.41	\\
ASAS$\_$J	073246-2047.5$^{a}$	&	1876.69086	&	0.8192500	&	8.40		&	ASAS$\_$J	200304-0256.0$^{b}$	&	2902.52596	&	0.4571957	&	10.60	\\
ASAS$\_$J	073516-4842.9	&	3462.58991	&	0.4310640	&	11.80		&	ASAS$\_$J	201125-0632.0	&	2144.61137	&	1.1884401	&	10.13	\\
ASAS$\_$J	075125+0454.6	&	3820.54204	&	0.7928613	&	10.16		&	ASAS$\_$J	201934-2726.8	&	4191.07671	&	0.4074481	&	12.32	\\
ASAS$\_$J	082205-6556.3	&	2618.71516	&	0.7419419	&	9.55		&	ASAS$\_$J	203141-5353.4	&	3705.55492	&	0.3519324	&	10.84	\\
ASAS$\_$J	082243+1927.0$^{a,b}$	&	3410.65967	&	0.2800459	&	11.26		&	ASAS$\_$J	203445-7237.0	&	4333.63344	&	0.4077932	&	10.61	\\
ASAS$\_$J	082704+0330.9$^{a}$	&	2977.79834	&	0.3278289	&	9.95		&	ASAS$\_$J	204625-1247.1	&	3554.77871	&	0.4275123	&	11.05	\\
ASAS$\_$J	083226-2910.1	&	2899.87530	&	0.6878800	&	10.56		&	ASAS$\_$J	204628-7157.0$^{a}$	&	3444.90678	&	0.7950000	&	8.62	\\
ASAS$\_$J	085335-7028.1	&	2619.75101	&	0.6389400	&	10.60		&	ASAS$\_$J	211902-0842.4	&	3581.81767	&	1.1746938	&	10.90	\\
ASAS$\_$J	091010+0344.6$^{a}$	&	3030.75836	&	0.4722758	&	11.09		&	ASAS$\_$J	212146-6544.2	&	4245.87043	&	0.6556112	&	9.66	\\
ASAS$\_$J	093547-1335.2	&	3672.87506	&	0.3510754	&	9.93		&	ASAS$\_$J	213209-3442.9	&	4295.83307	&	0.3766000	&	10.20	\\
ASAS$\_$J	093818-6755.4	&	3411.68022	&	0.3898983	&	10.26		&	ASAS$\_$J	214322+1442.5	&	3596.74982	&	0.5770898	&	11.27	\\
ASAS$\_$J	093838-5749.4	&	1977.58848	&	0.4382420	&	10.94		&	ASAS$\_$J	215035-2748.6$^{a,b}$	&	2974.53368	&	0.3738857	&	9.29	\\
ASAS$\_$J	100524-1416.3	&	2248.75486	&	0.4416701	&	11.51		&	ASAS$\_$J	215711+2240.2	&	3330.51545	&	0.4220225	&	9.56	\\
 \hline
\end{tabular}

\tablenotetext{a}{Systems that have already been studied. All of them are confirmed LMRs.}
\tablenotetext{b}{LMR systems included in the catalog of \cite{2022JApA...43...94W} that are potential red nova progenitors.}

\end{table*}



%\appendix

%\section{Appendix information}


\bibliography{bibliography}{}



\end{document}